\documentclass[prc,showpacs,nofootinbib,reprint,floatfix,aps,amsfonts,amssymb,amsmath]{revtex4-1} 
 
\usepackage[utf8x]{inputenc}

\usepackage{graphicx} 
\graphicspath{{figures/}}

\usepackage{bm}			
\usepackage{array} 
\usepackage{paralist} 
\usepackage{verbatim}
\usepackage{marginnote}

\usepackage{color}
\usepackage{subfigure}
\usepackage{float}

\newcommand{\bra}[1]{\ensuremath{\langle#1|}}
\newcommand{\ket}[1]{\ensuremath{|#1\rangle}}

\newcommand{\chieft}{\ensuremath{\chi}EFT}
\newcommand{\comments}[1]{}

\newcommand{\mnote}[1]{{\marginpar{\color{blue}\linespread{.7}\raggedright\footnotesize {#1}}}}
\newcommand{\bnote}[1]{\textbf{\color{red}\normalsize #1}}
\newcommand{\fbnote}[1]{\textbf{\color{green}\normalsize #1}}
\renewcommand{\mnote}[1]{}
\renewcommand{\bnote}[1]{}
\renewcommand{\fbnote}[1]{}

\newcommand{\vlowk}{V_{{\rm low}\,k}}
\newcommand{\oneSzero}{$^1$S$_0$}
\newcommand{\threeSone}{$^3$S$_1$}
\newcommand{\onePone}{$^1$P$_1$}
\newcommand{\infm}{\,\mbox{fm}^{-1}}
\newcommand{\avpot}{Argonne $v_{18}$}

\begin{document}

\title{Universality in similarity renormalization group evolved potential matrix elements and T-matrix equivalence}

\author{B. Dainton}
\affiliation{Department of Physics, The Ohio State University, Columbus, Ohio 43210, USA}

\author{R.\ J.\ Furnstahl}
\affiliation{Department of Physics, The Ohio State University, Columbus, Ohio 43210, USA}

\author{R.\ J.\ Perry}
\affiliation{Department of Physics, The Ohio State University, Columbus, Ohio 43210, USA}

\begin{abstract}
We examine how the universality of two-nucleon interactions evolved using similarity renormalization group (SRG) transformations correlates with T-matrix equivalence, with the 
ultimate goal of gaining insight into universality for three-nucleon forces. 
With sufficient running of the SRG flow equations, the low-energy matrix elements of different realistic potentials evolve to a universal form.  
Because these potentials are fit to low-energy data, they are (approximately)
phase equivalent only up to a certain energy, and we find universality in evolved potentials up to the corresponding momentum.  
More generally we find universality in \emph{local} energy regions, reflecting a local decoupling
by the SRG.
The further requirements for universality in evolved potential matrix elements are
explored using two simple alternative potentials.  
We see evidence that in addition to predicting the same observables, common long-range 
potentials (i.e., explicit pion physics) 
is required for universality in the potential matrix elements after SRG flow.  
In agreement with observations made previously for $\vlowk$ evolution, regions of universal potential matrix elements are restricted to where half-on-shell T-matrix equivalence holds.
\end{abstract}


\maketitle


\section{Introduction}

\begin{figure*}[tbh!]
	\includegraphics[width=2.2in]{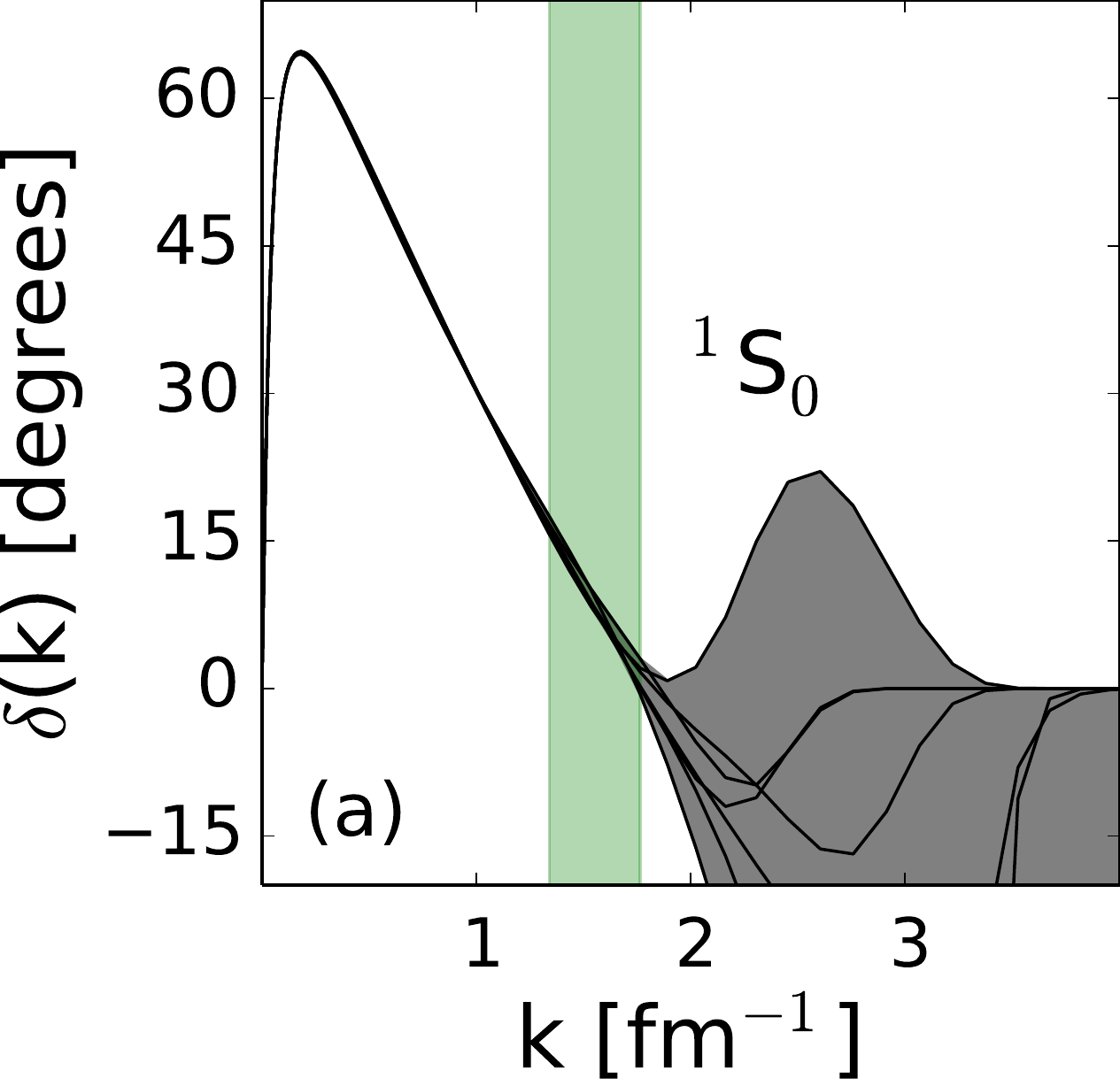}~~~%
	\includegraphics[width=2.2in]{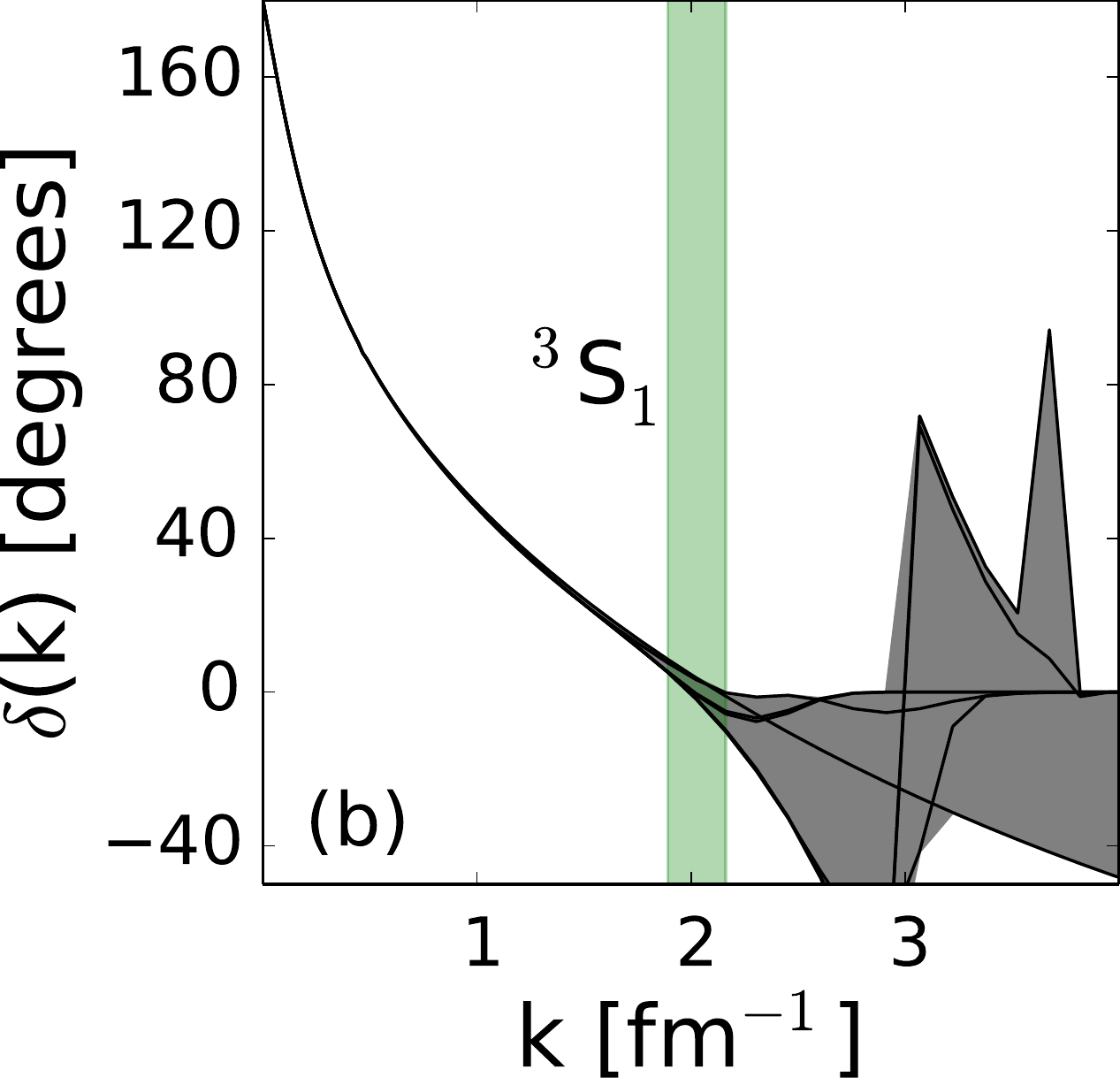}~~~%
	\includegraphics[width=2.2in]{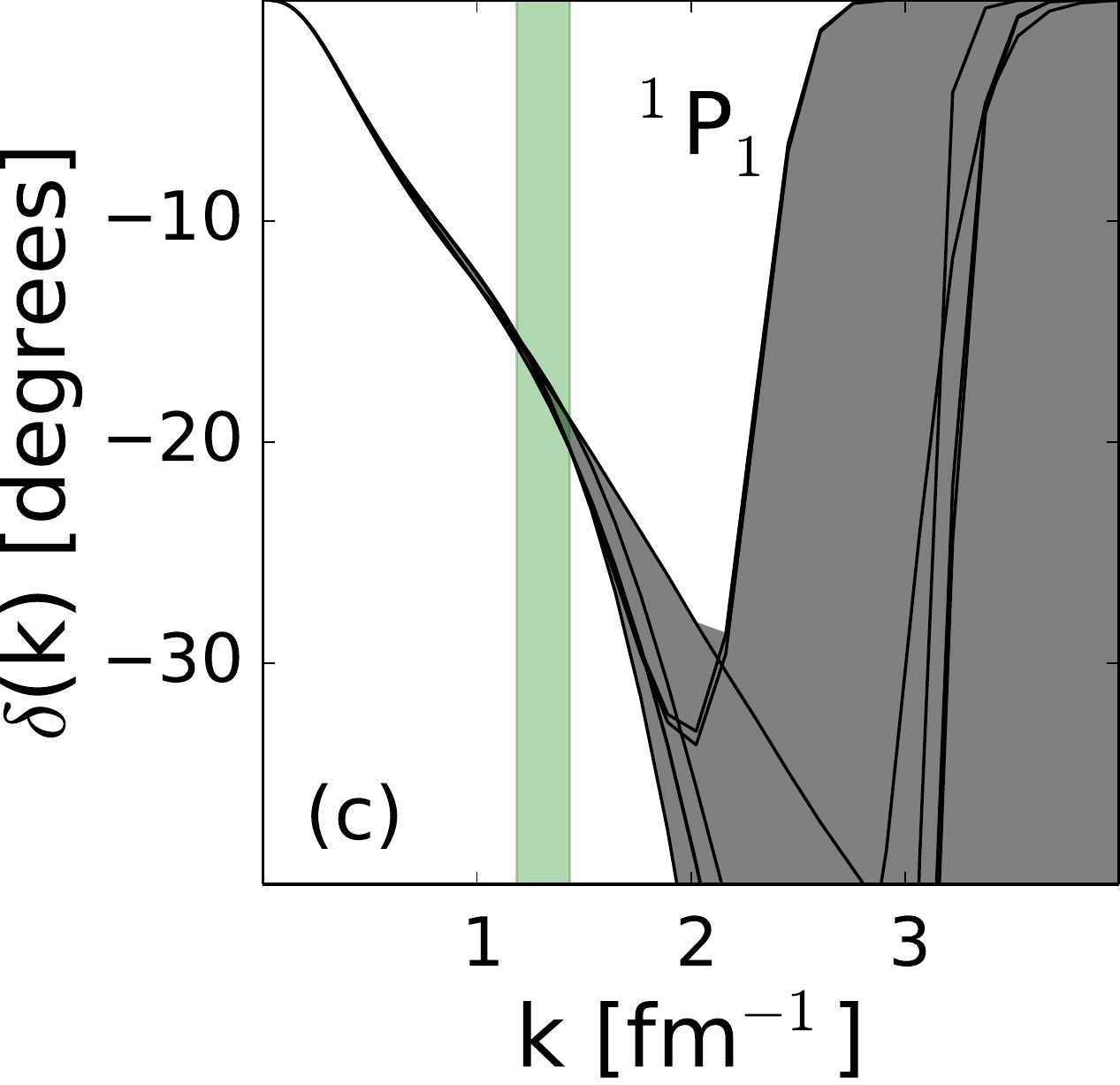}
   \vspace*{-.1in}
	\caption{(Color online) Phase shifts of various realistic potentials (see text) in the (a) \oneSzero,
	(b) \threeSone, and (c) \onePone\ partial waves.  The shaded regions show the range between
	the largest and smallest phase shifts. The vertical bands indicate the region
	where phase shift equivalence between the potentials ends, as defined
	by Eqs.~\eqref{eq:epsilon} and \eqref{eq:vertband}.
	\label{fig:ps_modern}}
\end{figure*}

A wide variety of realistic potentials are available 
for the low-energy nuclear two-body problem, including both
phenomenological interactions~\cite{av18orig,av18} and interactions   
motivated from systematic expansions, such as chiral effective field theory 
(\chieft)~\cite{EM500,EB,Epelbaum:2008ga}. 
The ability of these different potentials to reproduce the same low-energy observables
(e.g., see the phase shifts in Fig.~\ref{fig:ps_modern}) is one type of universality.
However, this universality does not generally extend to their Hamiltonian matrix elements,
which can vary drastically,  
reflecting the broad freedom to redefine interactions without changing S-matrix elements.

These realistic potentials lead to computational difficulties in most many-body
calculations, because requiring them to reproduce elastic phase shifts 
up to the pion-production threshold leads to strong coupling between low- and high-momentum 
matrix elements.
(The exceptions are \chieft\ potentials with low cutoffs and $J$-matrix-based
inverse scattering potentials~\cite{Shirokov:2003kk}.) 
For many-body methods using basis expansions, for example,
this coupling requires matrix sizes that become prohibitively large for accurate
microscopic calculations of nuclei.  
Thus we must face the problem of restrictions to smaller Hamiltonian matrices while maintaining the accuracy of predicted observables.

To address this problem, Lee-Suzuki transformations were applied in free space to integrate out high-energy degrees of freedom and soften an initial realistic potential, 
generating phase-equivalent ``low-momentum'' or ``$\vlowk$'' potentials~\cite{vlowkuniv,bognerfurnstahlschwenk}. 
This can be done in one step or using a renormalization group (RG) equation for the
potential~\cite{Bogner:2001jn}. 
Bogner and collaborators observed that a wide variety of realistic potentials have very similar low-momentum matrix elements after softening, which they termed the \textit{model independence} of $\vlowk$ potentials~\cite{Bogner:2001gq,vlowkuniv}.  
The diagonal $\vlowk$ potential matrix elements were found to match in regions of phase equivalence of the realistic potentials while the off-diagonal matrix elements matched in regions of half-on-shell (HOS) T-matrix equivalence~\cite{vlowkuniv}.  
They suggested that differences in the HOS T-matrix and thus the off-diagonal $\vlowk$ potential matrix elements occur because of different treatments of pion physics~\cite{vlowkuniv}.

Subsequently, similarity renormalization group (SRG) unitary transformations have been used 
to soften nuclear potentials while preserving 
observables~\cite{GlazekWilson,Wegner,Bogner:2006pc,bognerfurnstahlschwenk,Roth:2011vt,Furnstahl:2013oba}.  
Like $\vlowk$ transformations, the SRG decouples high-energy from low-energy physics, 
allowing one to truncate the matrices above some decoupling 
scale~\cite{decouplingSRG,SRGgeneral,bognerfurnstahlschwenk}.  
Further, the low-energy matrix elements of initial realistic potentials also
flow to the same form, but differ in detail from $\vlowk$ transformations. 
There is preliminary evidence that the SRG flow to common matrix elements extends to 
three-body forces~\cite{Hebeler:2012pr,Wendt:2013bla}, which are important ingredients for consistent treatments of nuclei 
with RG methods~\cite{Hammer:2012id,Furnstahl:2013oba}.

In analogy to the behavior of other Hamiltonians under RG transformations,
this model independence is naturally interpreted as a flow to 
universality in the evolved potential matrix elements.  
This form of universality can have powerful consequences if it can be understood
and exploited.
It suggests that for low-energy problems, a broad class of starting potentials that fits data will be equally effective after evolution~\cite{Timoteo:2011tt,Arriola:2013nja}. 
If realized for many-body forces, 
it may be possible to more easily construct accurate potentials (choosing operators based solely on the ease of use, then fitting constants to data), if they flow to a universal form after running the SRG.  

In applications of RG to \emph{local} quantum field theory, universality is a proven
tool.
When different theories are decomposed into relevant, marginal, and irrelevant interactions
according to their behavior under RG  flow, universality arises naturally among 
theories that share
the same relevant and marginal local interactions.
That is, if they differ only in the strength of their irrelevant couplings,
RG transformations reveal the universality as the RG flow rapidly eliminates
any irrelevant differences.
In nonrelativistic many-body theories that employ \emph{nonlocal} interactions, 
the possibility of universality in the form of phase equivalence is not a
surprise; inverse scattering theory and effective field theory imply that infinitely 
many potentials will yield the same low-energy results.
But the emergence of universal Hamiltonians (i.e., universal matrix elements) from RG flow
is not obvious without an operator classification that isolates irrelevant differences
between potentials.
To make progress in the absence of such a classification, we focus here on 
understanding the prerequisites for universality 
in SRG-evolved matrix elements, starting with two-body interactions.

We re-examine for the SRG the conclusions of Ref.~\cite{vlowkuniv} for $\vlowk$ potentials,
that the potentials must be phase equivalent up to a certain resolution scale but also have consistent, explicit handling of the long-range pion physics~\cite{bognerfurnstahlschwenk}.  
We use an inverse scattering separable potential (ISSP) to test if universality in potential matrix elements emerges at high energies and without explicit pion-exchange terms.  
The ISSP can reproduce all observables in the two-nucleon problem, and we will see explicitly that this is not enough for all low-momentum matrix elements to flow towards a universal form at finite cutoff.  
Also, when creating the ISSP we are free to choose a binding energy independent of the phase shifts, thus we can see the effect of differences in the binding energy on evolved 
low-momentum matrix elements.

To test the idea that the same explicit long-range treatment is required for flow to a universal form,
we introduce a second simple potential that is phase equivalent at low energies and includes explicit 
one-pion exchange (OPE).  We use the model proposed by Navarro P\'erez \textit{et al}.~\cite{Arriola,Perez:2013jpa}, which
combines the OPE potential with a sum of $\delta$-shell potentials.  This potential replaces the 
short-range physics with simple terms to be fit to phase shifts, while preserving the long-range force.

In section~\ref{sec:SRG}, we briefly review the SRG and decoupling and comment
on similarities with the $\vlowk$ RG.  
We discuss universality in matrix elements of modern realistic potentials in 
Section~\ref{sec:realistic}.  
The main focus will emerge in Section~\ref{sec:ISSP}, where we provide a working description of the ISSP formalism, examine universality in ISSP\rq{}s, and discuss the resulting insight into the prerequisites for universality.  
Section~\ref{sec:OPE} gives a description of the $\delta$-shell plus OPE potential
and examines the SRG flow of this potential to a universal form. 
We also comment on the SRG flow of the JISP16 potential. 
Finally, we conclude in Section~\ref{sec:conclusion} with a summary and the outlook
for the three-body problem.  
Although this is a study of universality only for two-nucleon interactions, it serves as a step toward more efficient handling of the three- and many-nucleon interactions.


\section{Similarity Renormalization Group} \label{sec:SRG}

The similarity renormalization group is a continuous series of infinitesimal unitary transformations acting on the Hamiltonian.  The simplest SRG transformations can
be expressed in differential form as a flow equation:
\begin{equation}
	\frac{dH_{s}}{ds} = [\eta_{s},H_{s}] = [[G_{s},H_{s}],H_{s}] \;,
	\label{eq:srgdiffeq}
\end{equation}
where s is a flow parameter~\cite{GlazekWilson,Wegner,bognerfurnstahlschwenk}.  
For most nuclear applications to date, the operator $G_{s}$ is the kinetic energy operator, 
denoted $T$.  
(We will refer to $G_s$ in this work as the SRG ``generator''.)
The most commonly used diagonalizing generator for non-nuclear applications is known as the
Wegner generator~\cite{Kehrein:2006}.  
It uses the diagonal of the Hamiltonian, $H_{s}^{d}$ instead of $T$ for $G_{s}$.  
Flows using the Wegner generator are indistinguishable from $T$ for the range of evolution in the present study but can differ drastically if the SRG cutoff becomes very low~\cite{GlazekPerry} or if a large-cutoff chiral potential is used~\cite{Wendt:2011qj}.

The goal of the SRG is to decouple high-energy from low-energy degrees of freedom 
in the Hamiltonian by driving far off-diagonal matrix elements to zero.  
Instead of $s$, we usually refer to the decoupling scale, $\lambda = s^{-\frac{1}{4}}$ 
for $T$ and $H_{s}^{d}$, where $\lambda$ is chosen to have the same units as momentum.  
In the SRG flow with the $T$ generator, 
the dominant term of Eq.~\eqref{eq:srgdiffeq} for far off-diagonal matrix elements  
is the term linear in the potential, $[[T,V_{s}],T]$, where $V_{s} \equiv H_{s} - T$.
If we keep just this term, the flow equation is immediately solved for these
matrix elements, yielding (with mass $m=1$)
\begin{equation}
	V_{s}(k,k') \simeq V_{s=0}(k,k') e^{-(\frac{k^{2}-k'^{2}}{\lambda^{2}})^{2}}
	\;. 
\end{equation}
Thus $\lambda^{2}$ is roughly the maximum difference between
kinetic energies of nonzero matrix elements.
Once the Hamiltonian is sufficiently evolved to exhibit decoupling, low-energy observables 
can be obtained from a truncated Hamiltonian~\cite{decouplingSRG} or one finds naturally
that a smaller expansion basis is needed for a desired degree of convergence.  

A \emph{nondiagonalizing} alternative for $G(s)$ is the block generator, 
$H_{s}^{bd}$ defined in Ref~\cite{BlockDiag}.  
$H_{s}^{bd}$ matrix elements are the block diagonal elements of the evolved Hamiltonian $H_s$, 
separated at a fixed chosen cutoff parameter $\Lambda$. 
(That is, the generator $H_{s}^{bd}$ in a momentum basis 
is obtained from $H_{s}(k,k')$ by setting to zero the matrix elements where $k<\Lambda$
and $k'>\Lambda$ or $k>\Lambda$ and $k'<\Lambda$.) 
This is the same pattern of decoupling achieved with $\vlowk$ Lee-Suziki 
transformations~\cite{bognerfurnstahlschwenk,Arriola,vlowkuniv}.  
In fact, the $\vlowk$ and SRG block diagonal transformations have been shown to result in very similar 
Hamiltonians for the lower energy block if the SRG transformation is run to
$\lambda \ll \Lambda$~\cite{BlockDiag}.  
For $H_{s}^{bd}$, $\lambda = s^{-\frac{1}{2}}$ and represents the maximum difference in
energy for coupling between the blocks above and below $\Lambda$.
It has been shown that SRG with the $T$ generator~\cite{SRGgeneral} and $\vlowk$~\cite{bognerfurnstahlschwenk,Arriola,vlowkuniv} each drive realistic potentials to separate low-energy universal forms, and we will show that $H_{s}^{bd}$ also drives potential matrix elements to a different universal form.  
Because a nondiagonalizing transformation exhibits universality in low-energy potential matrix elements, universality cannot simply be a consequence of the generator $\eta_{s} = [T,H_{s}]$ driving potentials toward the diagonal.


\section{Modern Realistic Potentials} \label{sec:realistic}
	
We have chosen a representative phenomenological potential and a set of \chieft\
potentials to evolve and examine in various partial waves.  
The phenomenological potential is \avpot (AV18), which employs basis operators in position representation and fits the coupling constants to elastic scattering data~\cite{av18orig,av18}.  
We use the N$^{3}$LO \chieft\  potential from Entem and Machleidt with a cutoff of 
500 MeV~\cite{EM500} and then  
five N$^{3}$LO \chieft\  potentials with various cutoffs from Epelbaum \textit{et al}.~\cite{EB}.  
These \chieft\  potentials have different regularization and phase shift fitting schemes,
which creates differences in the matrix elements of the potentials. 

\begin{figure*}[htp!]
	\includegraphics[width=2.2in]{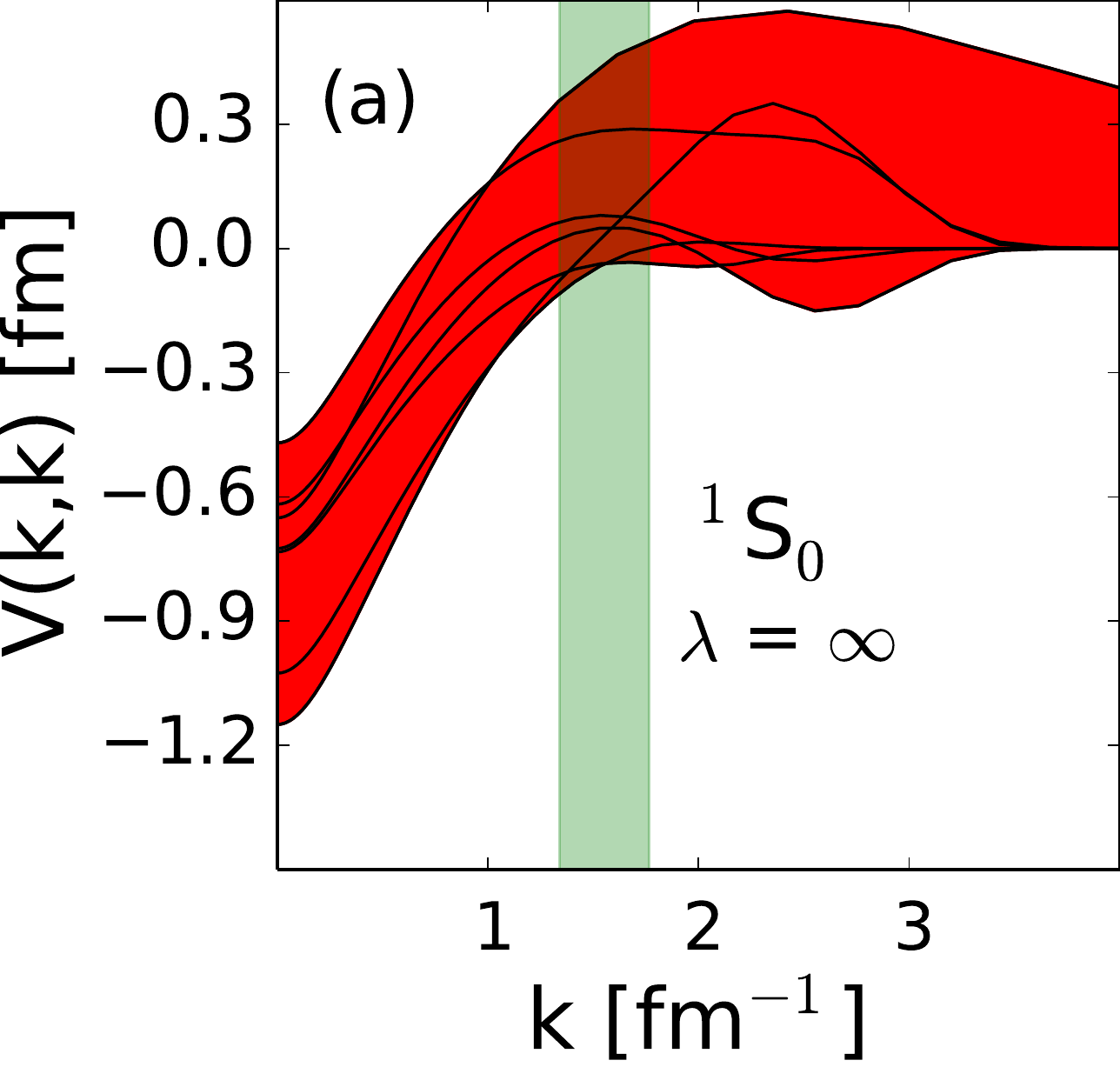}~~~%
	\includegraphics[width=2.2in]{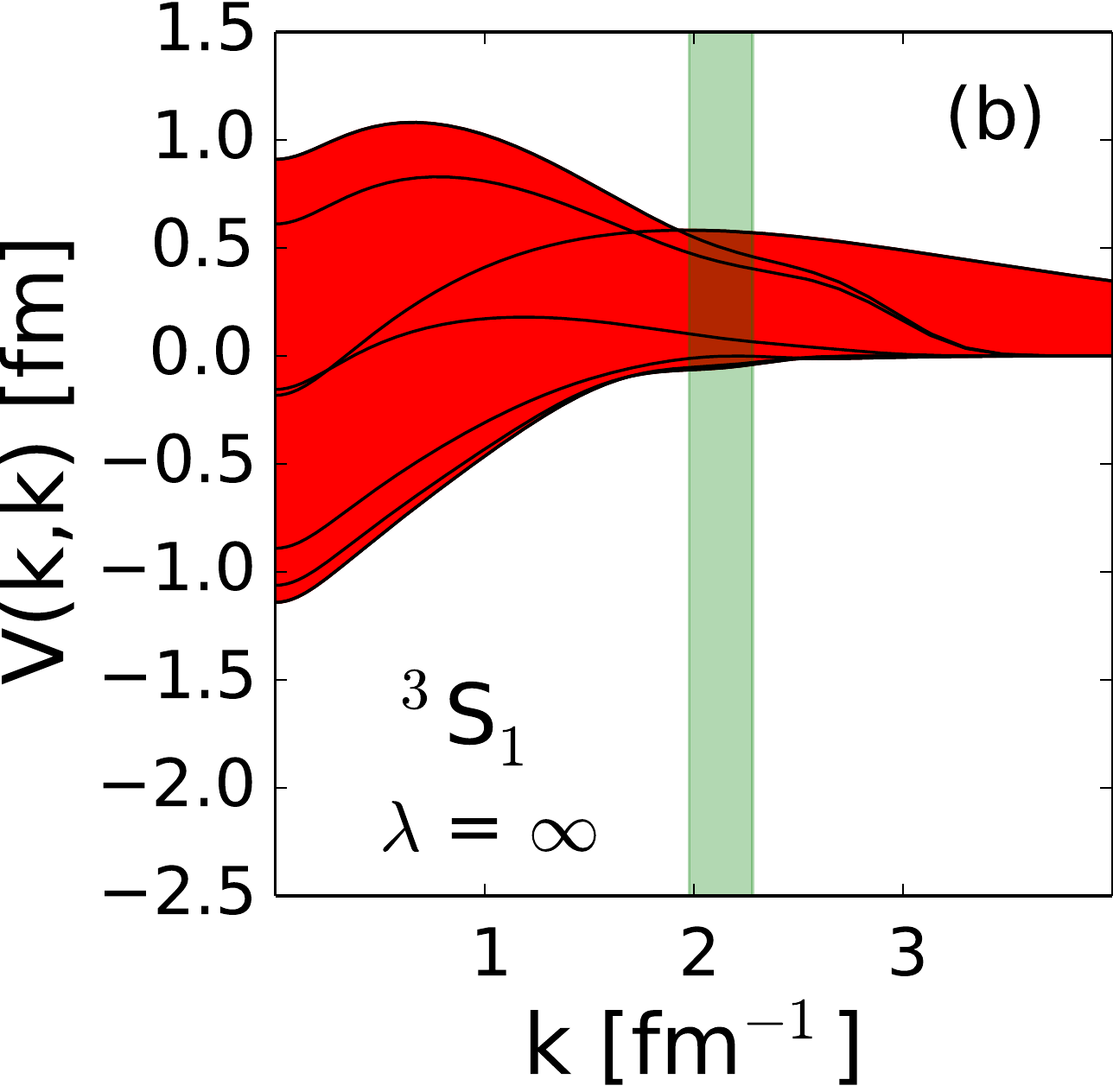}~~~%
	\includegraphics[width=2.1in]{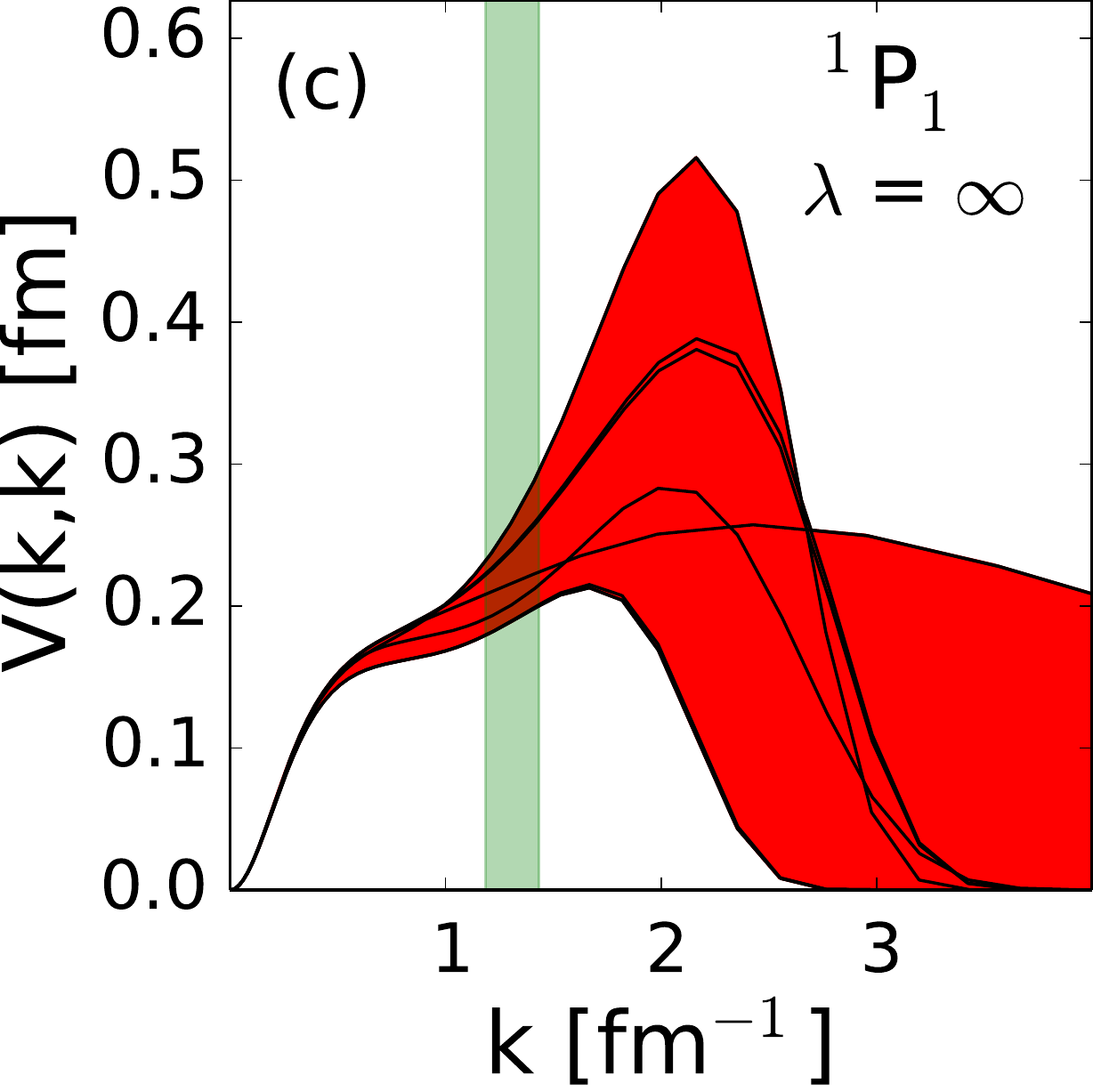}
   \vspace*{-.1in}
	\caption{(Color online)  Diagonal matrix elements $V(k,k)$ of various unevolved realistic potentials (see text) in the (a) \oneSzero, (b) \threeSone, and (c) \onePone\ partial waves.
	The shaded regions show the range of values and the vertical bands are from
	Fig.~\ref{fig:ps_modern}. 
	\label{fig:diag_modern_inf}}
   \vspace*{.1in}
	\includegraphics[width=2.2in]{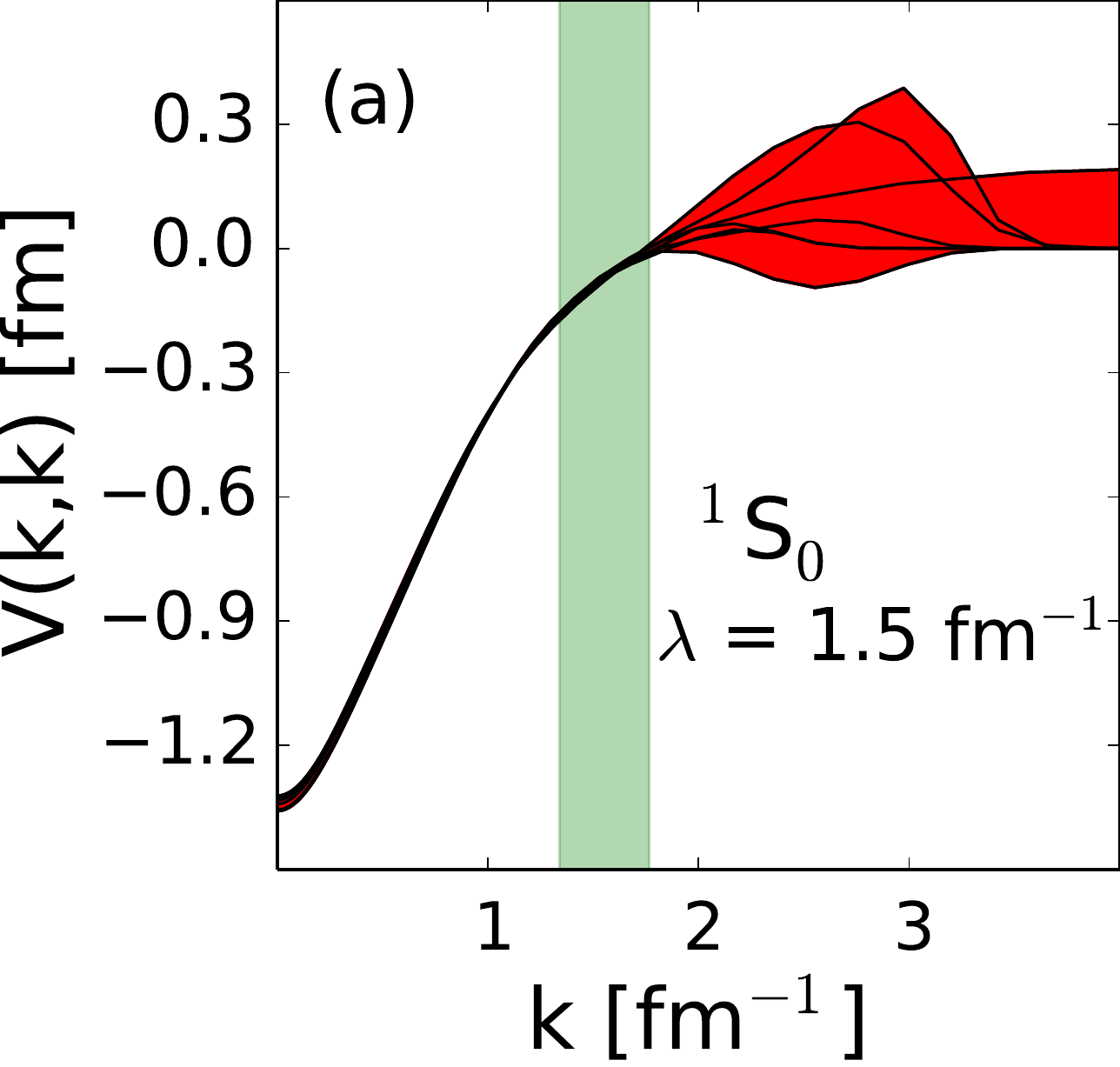}~~~%
	\includegraphics[width=2.2in]{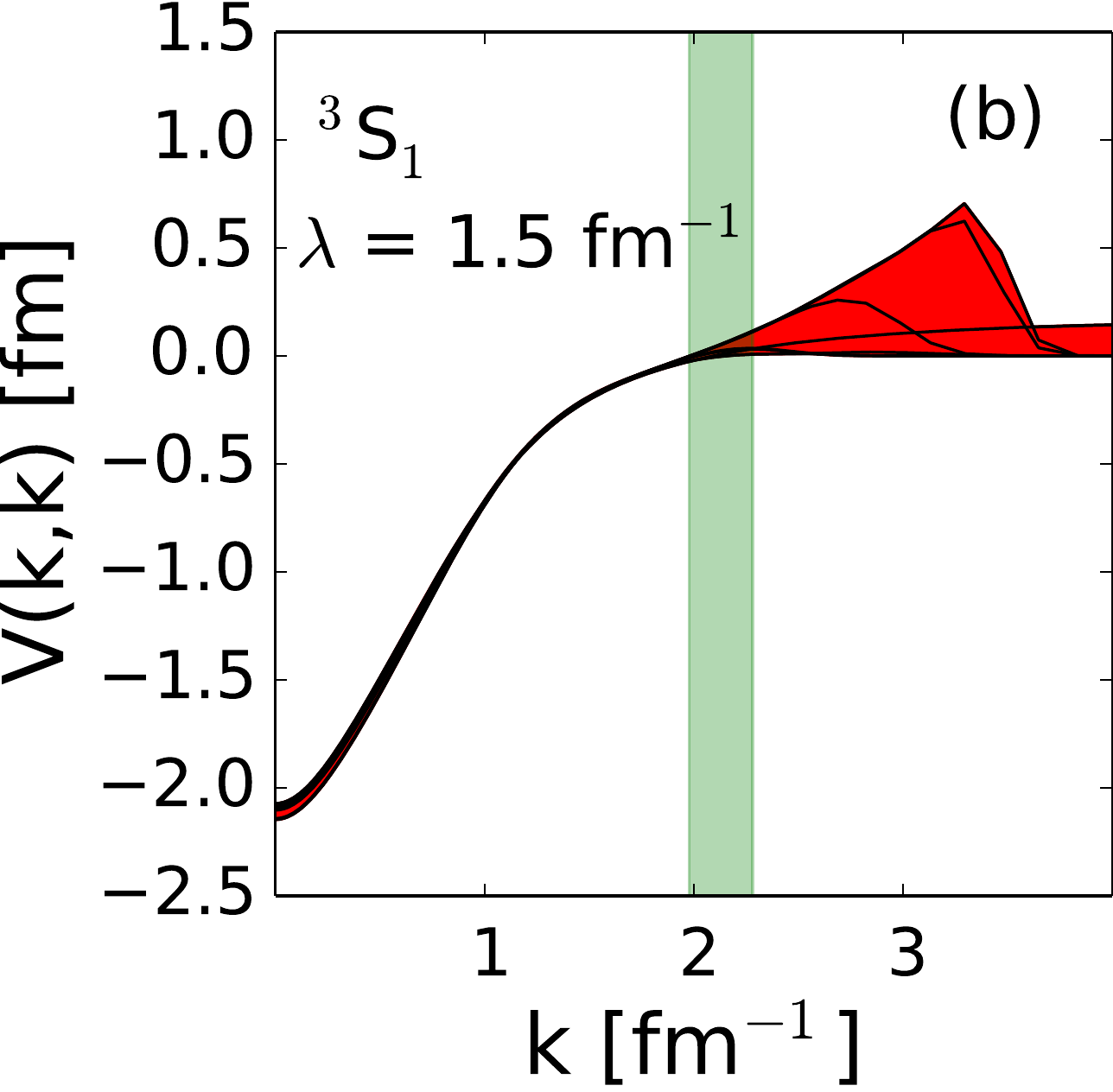}~~~%
	\includegraphics[width=2.1in]{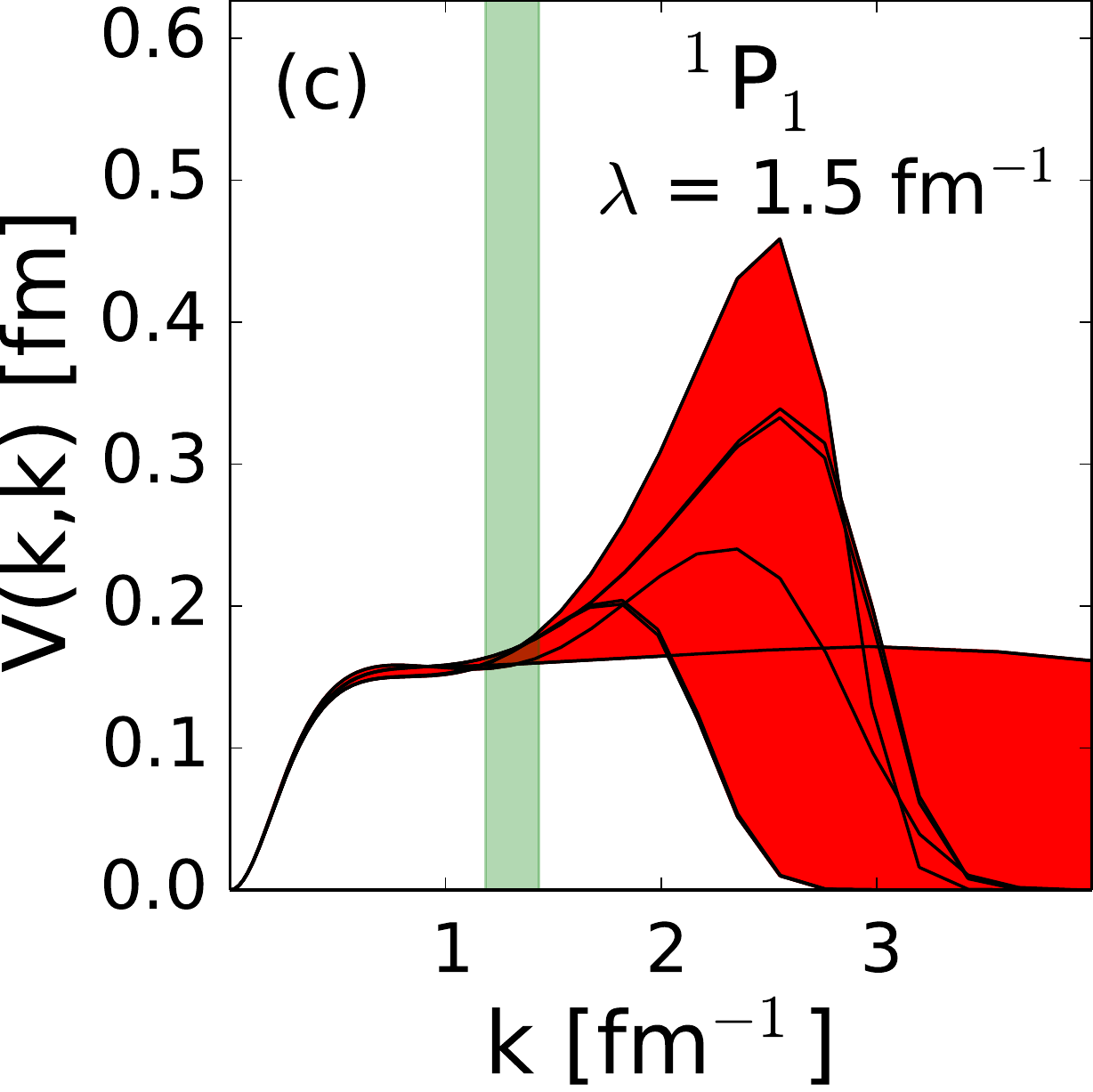}
   \vspace*{-.1in}
	\caption{(Color online)  Diagonal matrix elements of various realistic potentials
	 in the (a) \oneSzero, (b) \threeSone, and (c) \onePone\  partial waves evolved 
	by the SRG to $\lambda$ = 1.5 fm$^{-1}$. 
	The shaded regions show the range of values and the vertical bands are from
	Fig.~\ref{fig:ps_modern}. 
	\label{fig:diag_modern_15}}
   \vspace*{.1in}
	\includegraphics[width=6.9in]{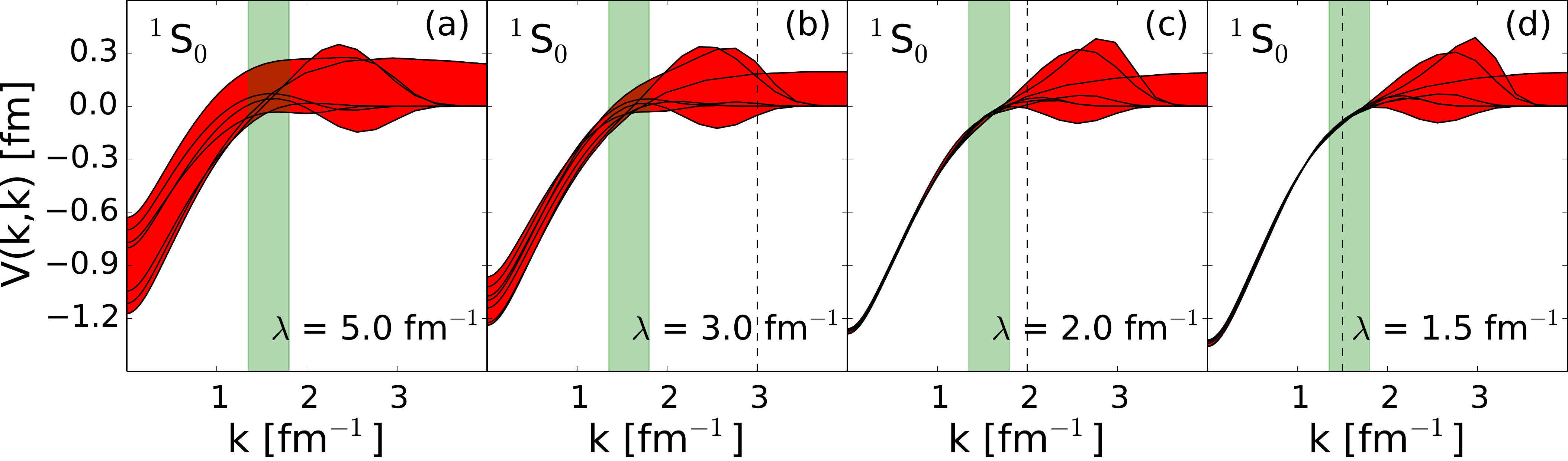}
   \vspace*{-.1in}
	\caption{(Color online) Diagonal matrix elements of various realistic potentials in
	the \oneSzero\ partial wave 
	evolved by the SRG to $\lambda$ = (a) 5.0 fm$^{-1}$, (b) 3.0 fm$^{-1}$, (c) 2.0 fm$^{-1}$, (d) 1.5 fm$^{-1}$ (marked by the vertical dashed line). 
	The shaded regions show the range of values and the vertical bands are from
	Fig.~\ref{fig:ps_modern}. 
	\label{fig:diag_modern_multi_cut}}
\end{figure*}

From Fig.~\ref{fig:diag_modern_inf} one can see that the diagonals of the 
initial potentials
in momentum representation are quite different (the differences are particularly
evident in lower partial waves, so we focus on those).
In making these comparisons, we do not single out individual potentials but use a
shaded region to highlight the range of matrix element variation.
As advertised, after evolution the matrix elements collapse at low momentum
to a universal dependence on momentum (the result at fixed $\lambda = 1.5\infm$
is shown in Fig.~\ref{fig:diag_modern_15}).  
This feature is not restricted to the diagonal elements; low-energy off-diagonal matrix
elements of the potentials also evolve to universal values (see Fig.~\ref{fig:dsuniv} below).  
At higher momentum, the potential matrix elements deviate.

\begin{figure*}[tbph!]
	\subfigure{\includegraphics[width=2.2in]{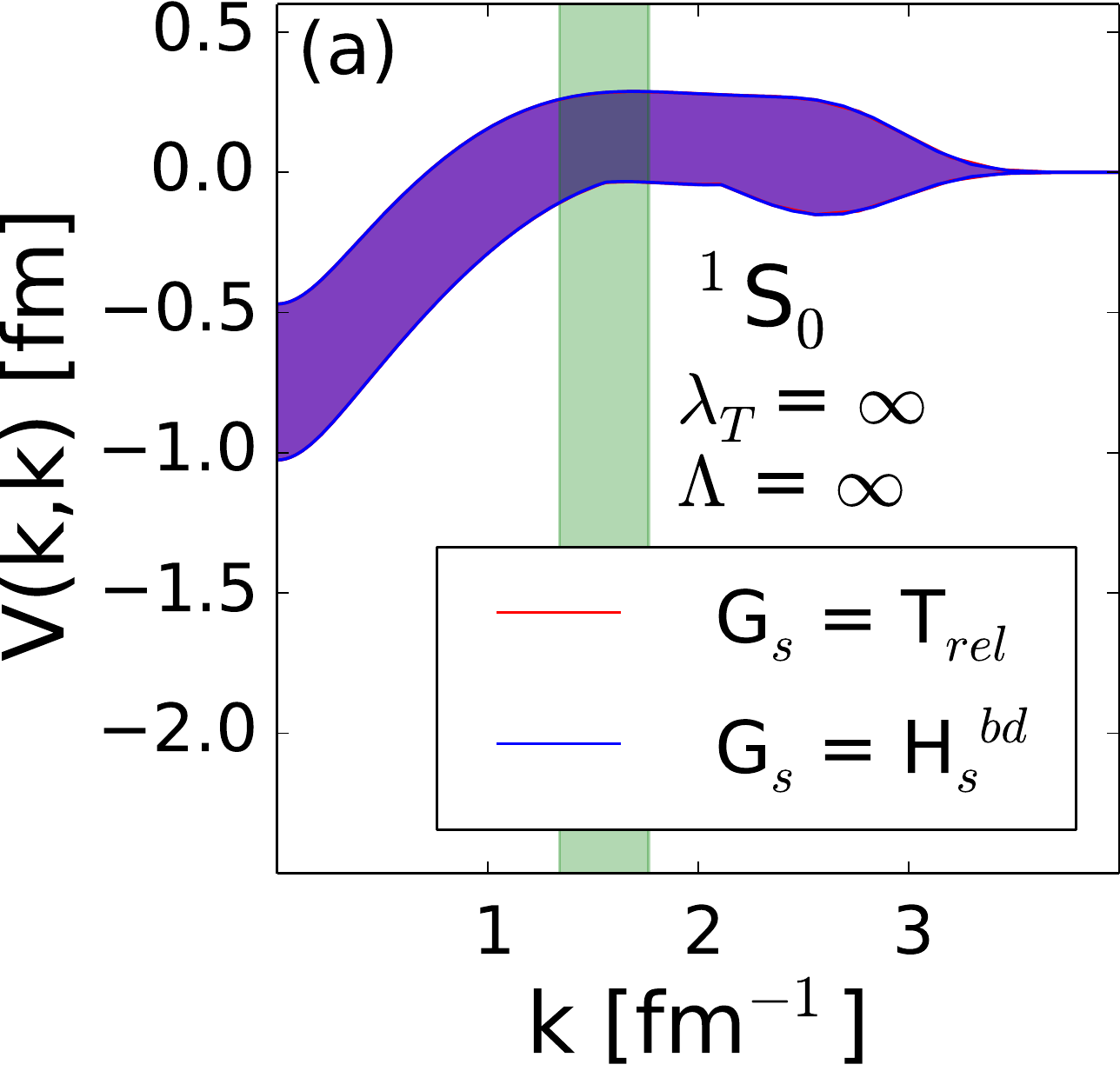}}~~~
	\subfigure{\includegraphics[width=2.2in]{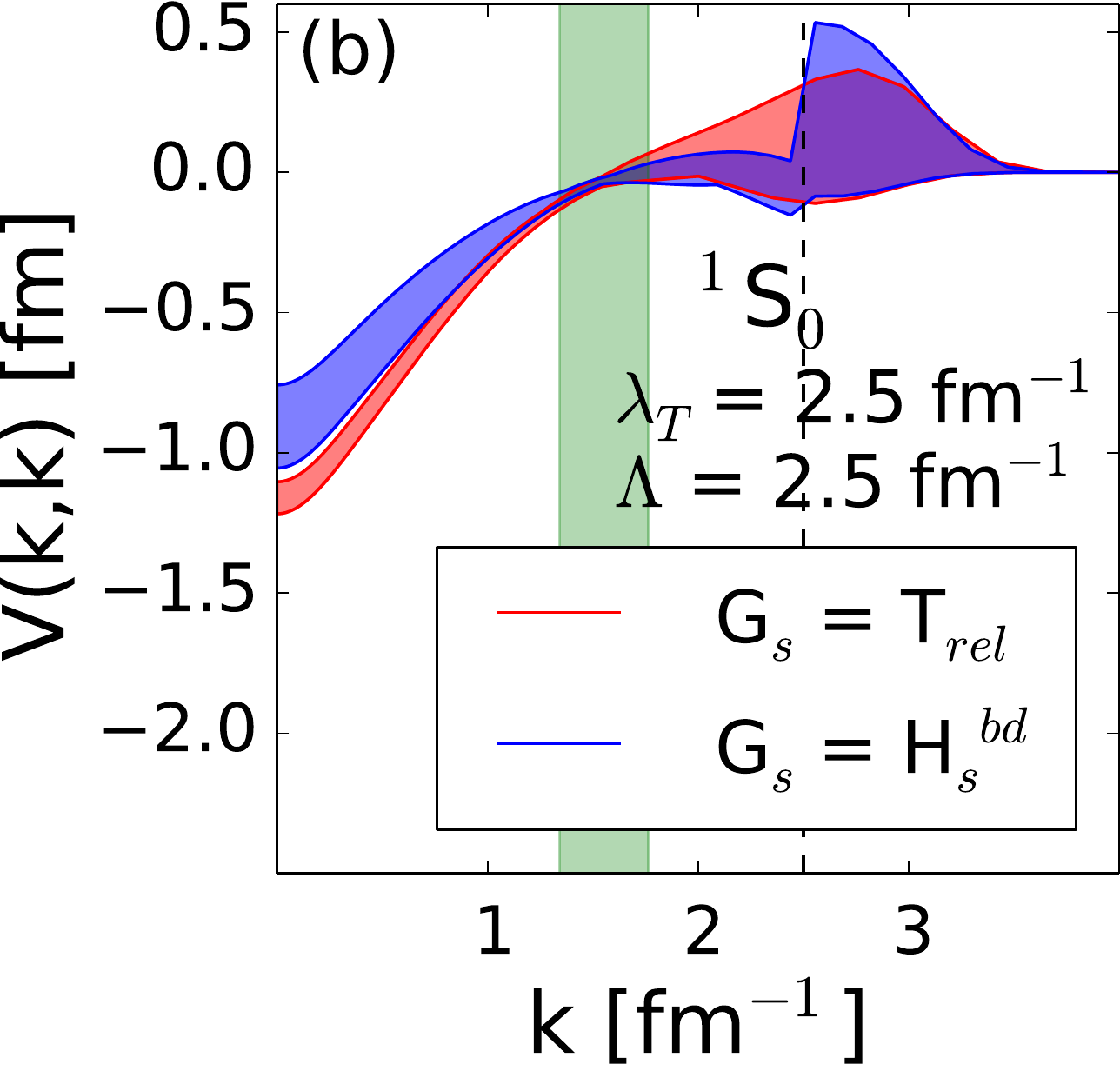}}~~~
	\subfigure{\includegraphics[width=2.2in]{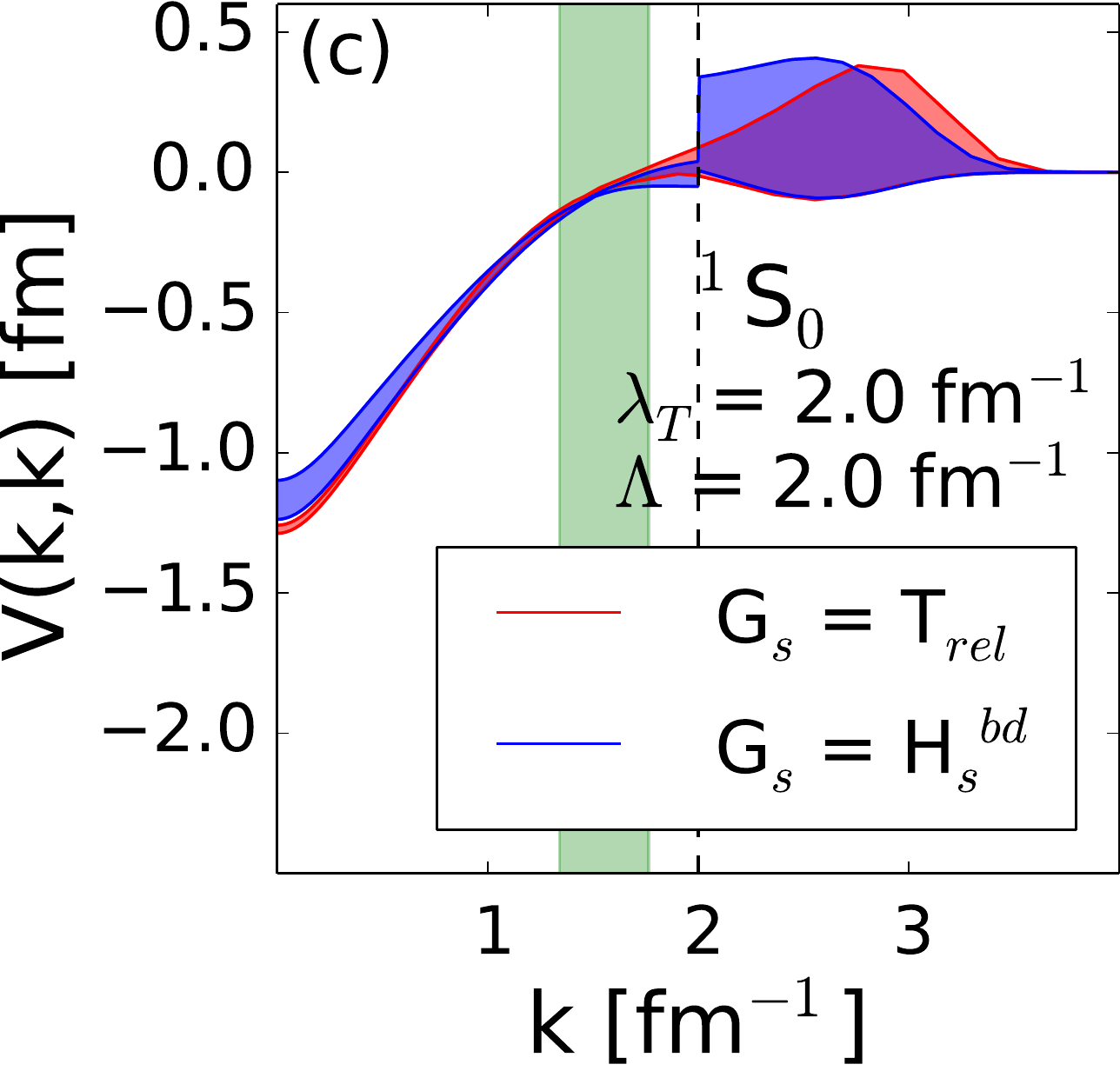}}
   \vspace*{.1in}
	\subfigure{\includegraphics[width=2.2in]{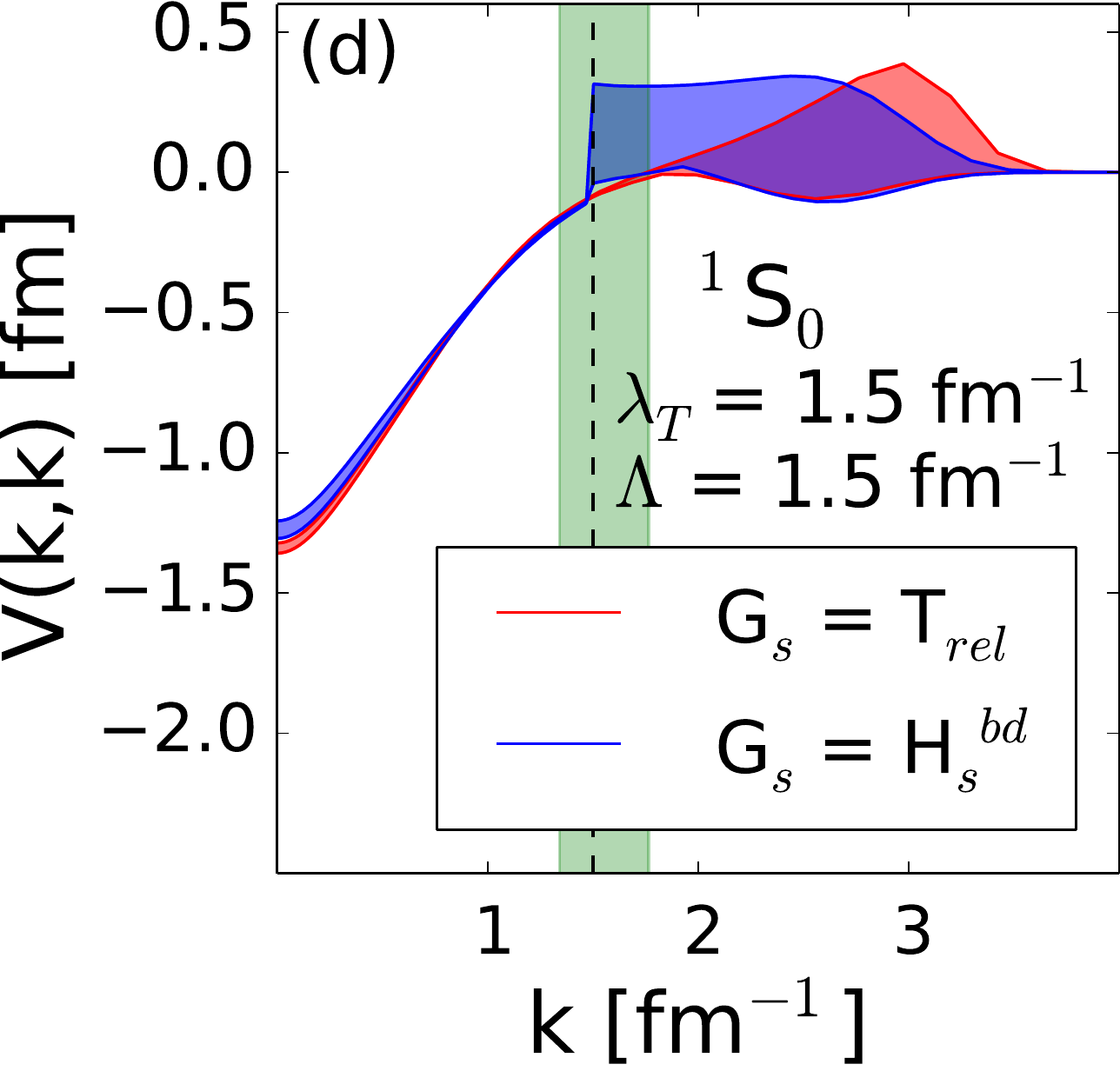}}~~~
	\subfigure{\includegraphics[width=2.2in]{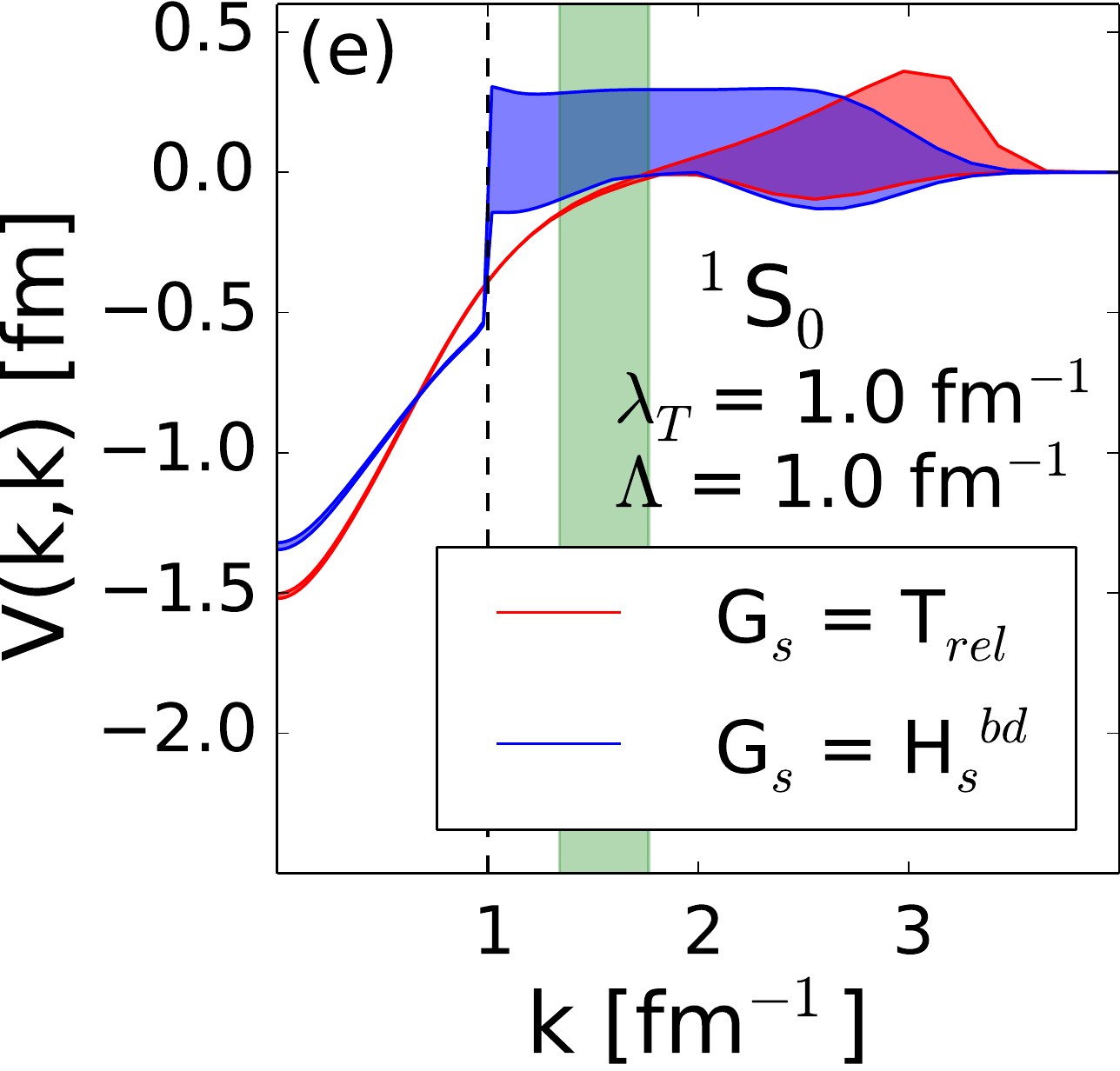}}~~~
	\subfigure{\includegraphics[width=2.2in]{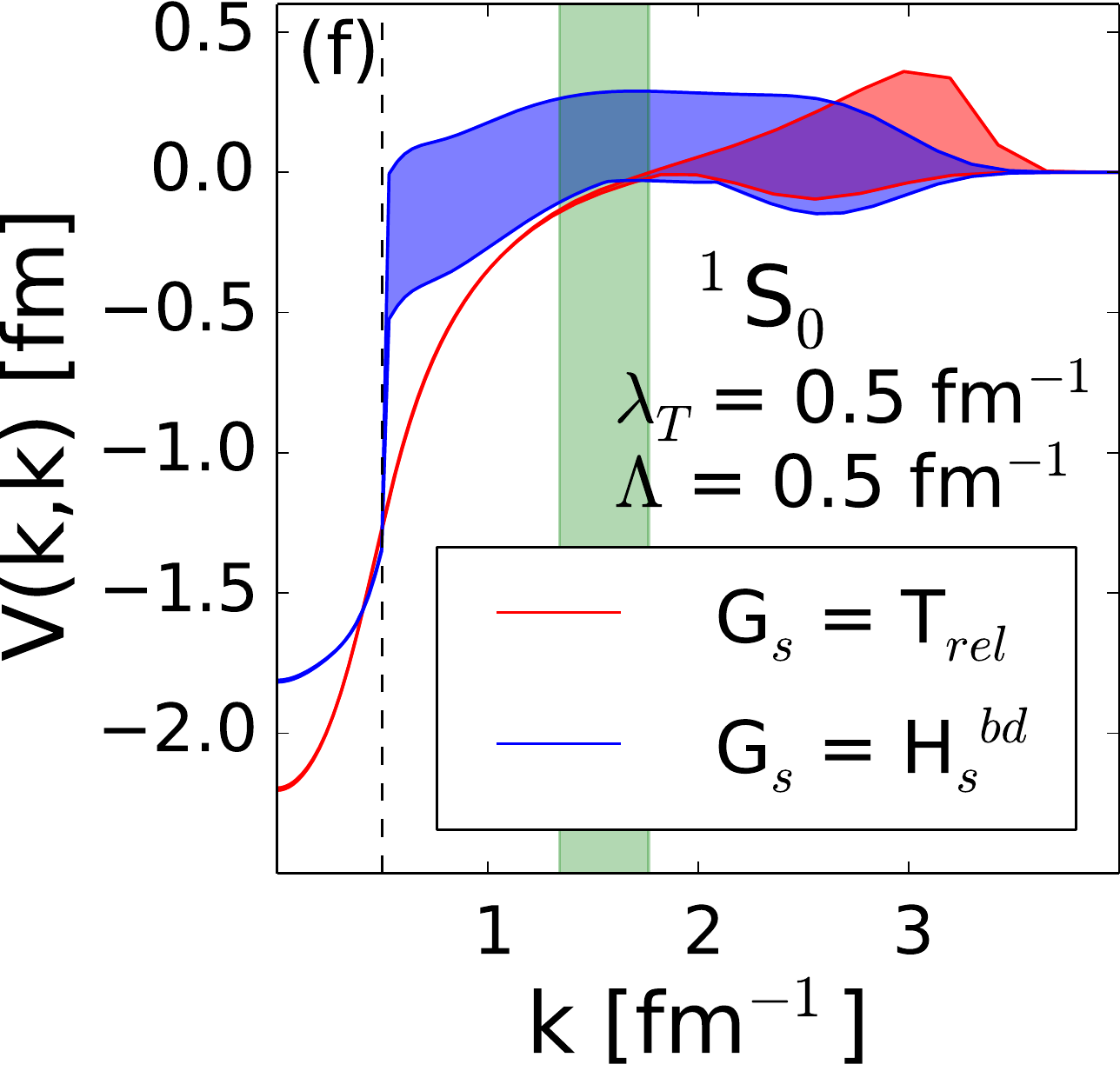}}
   \vspace*{-.1in}
	\caption{(Color online) The spread of diagonal matrix elements of various \chieft\ potentials (see text) 
	in the \oneSzero\ 
	partial wave are shown as shaded regions for the unevolved potential and then after
	evolution to $\lambda_{T}$ = 2.5, 2.0, 1.5, 1.0, and 0.5 fm$^{-1}$ with the $T$ generator 
	(red or light gray).
	These are compared to the spread of the corresponding matrix elements for the $H_{s}^{bd}$ generator
	with $\Lambda$ = 2.5, 2.0, 1.5, 1.0, and 0.5 fm$^{-1}$fm$^{-1}$, all 
	evolved to $\lambda$ = 0.5 fm$^{-1}$ (blue or medium gray).
	The vertical bands are from Fig.~\ref{fig:ps_modern} and the vertical dashed lines mark
	$\lambda_{T}$ or $\Lambda$.  
	\label{fig:blockdiag}}
\end{figure*}

Following Ref.~\cite{vlowkuniv}, we compare phase shift and matrix element deviations
to identify the correlations between phase equivalence and matrix element
universality.
In Fig.~\ref{fig:ps_modern}, we have identified vertical bands within which the phase shift equivalence among the various potentials ends and significant deviation begins.
While identifying an exact point marking this deviation will be somewhat arbitrary,
we can roughly choose a normalized width description that is consistent with 
visual assessments of the phase shift plots. 
In particular, for each partial wave, the vertical band represents the region characterized by:
\begin{equation}
   0.03 < \epsilon(k) < 0.1 \;,
   \label{eq:epsilon}
\end{equation}
where
\begin{eqnarray}
	\epsilon(k) &\equiv& \frac{\delta_{\rm high}(k)-\delta_{\rm low}(k)}{\Delta} 
	  \;. 
   \label{eq:vertband}
\end{eqnarray} 
The numerator is the range of phase shifts at a fixed $k$ while $\Delta$ is the range of phase 
shifts for the entire universality region. 
Our studies imply that the precise definition of $\epsilon$ is not important; as long as it consistently identifies
the regions where phase equivalence ends it can be used to consistently compare to the regions where 
the universality of matrix elements end.

Comparing Figs.~\ref{fig:diag_modern_inf} and \ref{fig:diag_modern_15}, we see that
while diagonal matrix elements of the initial potentials differ significantly in the region
where phase equivalence ends, this same region corresponds to where the matrix
elements have collapsed to universal values by $\lambda = 1.5\infm$.  
This suggests the hypothesis that \emph{a prerequisite for 
matrix element universality is phase equivalence}.  Namely, if there are 
\emph{local} regions in energy
in which potentials are not phase equivalent, then there is no universality in those regions
(this is tested further in Section~\ref{sec:ISSP}).  
Examining the diagonals of the potentials more closely, we observe that for the \oneSzero\
and \threeSone\ channels, the lowest matrix elements are not exactly the same.  
This may be a consequence of not evolving $\lambda$ further.
From the $T$ generator curves in Fig.~\ref{fig:blockdiag}, we can see that the slight width of the band 
decreases as we evolve chiral potentials to $\lambda = 0.5 \infm$.
Also, as we will see below, differences in the binding energy of the deuteron play 
an important role in the low-energy matrix elements of the \threeSone\ potential.

How low must $\lambda$ be before we see universality? 
Figure~\ref{fig:diag_modern_multi_cut} shows the diagonals of the \oneSzero\ potential evolved
to four different $\lambda$ values.  
The vertical bands correspond to the same region where phase equivalence ends for the
\oneSzero\ channel as in Fig.~\ref{fig:ps_modern}, while the vertical dashed line shows the 
value of $\lambda$.  
We see in this partial wave (and in others not shown as well) that universality in the matrix
elements does not occur until $\lambda$ approaches the vertical band.
A natural hypothesis is that the matrix elements will not fully collapse to universal form 
until $\lambda$ reaches the region of phase equivalence.  
There may be an intrinsic low-energy scale common to each of these
potentials that determines at which $\lambda$ universality in potential matrix elements will
appear.  A possibility is that this scale is a consequence of explicit treatment of pion physics in each of the modern realistic potentials.  
To test the latter explanation, a potential with phase equivalence at much higher momenta 
and no explicit pion physics is required, which we consider in the next section.

As described earlier, the block-diagonalizing generator H$_{s}^{bd}$ will drive the potential
matrix elements to a different universal form than $T$.  
This is illustrated in Fig.~\ref{fig:blockdiag} with a set of \chieft\ potentials in
the \oneSzero\ channel.
When evolved to $\lambda \leq 2\infm$ with the $T$ generator, the universal form of diagonal potential matrix elements emerges over the full region of phase equivalence.  
For the block diagonal generator with $\Lambda \leq 2\infm$, however, only diagonal matrix elements below $\Lambda$ become universal and with a different
flow than the matrix elements evolved with $T$.  
The universality is only up to $\Lambda$ because this SRG only decouples one block from
the other, so matrix elements at momenta above $\Lambda$ still couple to matrix elements 
in phase inequivalent regions and therefore do not collapse to a universal form. 
(Note that in the $\vlowk$ RG, the higher block is set to zero.) 
We will discuss only $G_s = T$ in the rest of this study but emphasize that the ideas about universality apply to both generators, although only in the low-momentum block for the 
H$_{s}^{bd}$ SRG.

The region of phase equivalence for the realistic potentials is limited by the energies 
to which they can be fit to elastic scattering phase shifts.  
Because of this, if we wish to investigate different regions of universality, we must use a method that can \lq{}fit\rq{} the phase shifts in a controlled range of energies.  
One of the simplest approaches is solving the inverse scattering problem with a separable potential, which we consider in the next section.


\section{Separable Inverse Scattering Potential} \label{sec:ISSP}

Instead of fitting coupling constants for predetermined operators to the phase shifts, 
an inverse scattering procedure constructs a potential directly from the phase shifts.
Separability is just a constraint to define a unique potential, chosen here due to its 
simplicity.
For instance, when solving the Lippmann–Schwinger equation, a separable potential 
reduces the problem of solving an integral equation to simply evaluating an integral.  
The three-body Faddeev equations also simplify for a separable potential, as one of the integrals 
over internal momenta becomes trivial. 
A key feature of the ISSP for this study is that the potential is entirely created from 
the phase shifts and binding energy of the deuteron; no \emph{explicit} pion exchange 
or other physics is imposed.  
This allows us to determine whether or not universality requires extra physics, 
such as explicit long-range pion terms or other phenomenological considerations.
We start with a brief summary of the inverse scattering separable potential for two
nucleons. 

\subsection{Formalism}

The form of a rank-$n$ separable potential is:
\begin{equation}
		V = \sum_{i,j=0}^{n-1} \ket{\nu_{i}} \Lambda_{ij} \bra{\nu_{j}} \;.
\end{equation}
For our purposes a rank-1 separable potential will be sufficient, but future studies
may benefit from a higher-rank potential.
A rank-1 potential in momentum representation takes the form:
	\begin{equation}
		V(k,k\rq{})=\sigma \nu(k) \nu (k\rq{}) \;,
		\label{eq:separable}
	\end{equation}
where $\sigma$ is simply $\pm 1$.	
Details of the rank-1 separable inverse scattering problem are well 
documented~\cite{BrownJackson,Kwong}; here, we simply state the main results, 
some limitations, and how to work around the limitations.   
The solution to the separable inverse scattering problem 
is~\cite{BrownJackson}:
\begin{eqnarray}
		\sigma \nu^{2}(k) &=& - \frac{k^{2} - k_{b}^{2}}{k^{2}} 
		   \frac{\sin(\delta(k))}{k} e^{-\Delta(k)} \;,\\
		\Delta(k) &=& \frac{1}{\pi} P \int_{0}^{\infty} \frac{dk\rq{} \delta(k\rq{})}{k\rq{} - k} \;,\\
		E_{b} &=& \frac{\hbar^{2} k_{b}^{2}}{2 m}  \;,
\end{eqnarray}
where $k_{b}$ is zero if there is no bound state and equal to the binding momentum 
for a single bound state with binding energy $E_b$
(for a rank-1 separable potential there can be at most one bound state).
	
Once $\nu(k)$ is determined, the entire potential is known from Eq.~\eqref{eq:separable}.  
The binding energy $E_{b}$ can be tuned independently of the phase shifts.  
A limitation of rank-1 separable potentials is that if the phase shift as a function of 
momentum cross zero, then so too must the potential, and a rank-1 ISSP 
as defined thus far can never change signs if $\nu$ is real.
This point is clear from
Eq.~\eqref{eqzeroes}, which follows from the Lippmann-Schwinger equation for a 
separable potential (with standing wave boundary conditions):
	\begin{equation}\label{eqzeroes}
		\frac{1}{k} \tan(\delta_{l}(k)) = 
		-\frac{ V_{l}(k,k)}{1+\frac{2}{\pi}\mathcal{P}\int 
		    \frac{\textstyle dp \; p^{2} V_{l}(p,p)}{\textstyle p^{2}-k^{2}}} \;.
	\end{equation}
A zero-crossing in $\delta(k)$ corresponds to a zero-crossing on the right side
of this equation, which can only be achieved by the numerator crossing zero if 
the denominator remains finite.

Because some of the phase shifts for nucleon-nucleon partial waves exhibit zero crossings, 
we need an inverse scattering potential that allows this feature.  
We can still use the same rank-1 formalism, however, if we split the problem into two energy
regimes, above and below the zero crossing~\cite{Kwong}.  Then we can define:
	\begin{eqnarray}
		\delta_{<}(k) &\equiv& \delta(k) \theta(k_{0} - k) \;, \\
		\delta_{>}(k) &\equiv& \delta(k) \theta(k - k_{0}) \;, \\
		V(k,k\rq{}) &=&  V_{<}(k,k\rq{}) + V_{>}(k,k\rq{}) \;,
	\end{eqnarray}
and determine $V_{<}$ and $V_{>}$ separately using the rank-1 formalism 
with $\delta_{<}$ and $\delta_{>}$ as input, respectively.  We have confirmed
numerically that potentials created with this prescription
accurately reproduce the input phase shifts.

\begin{figure*}[htp!]
	{\includegraphics[width=.32\textwidth]{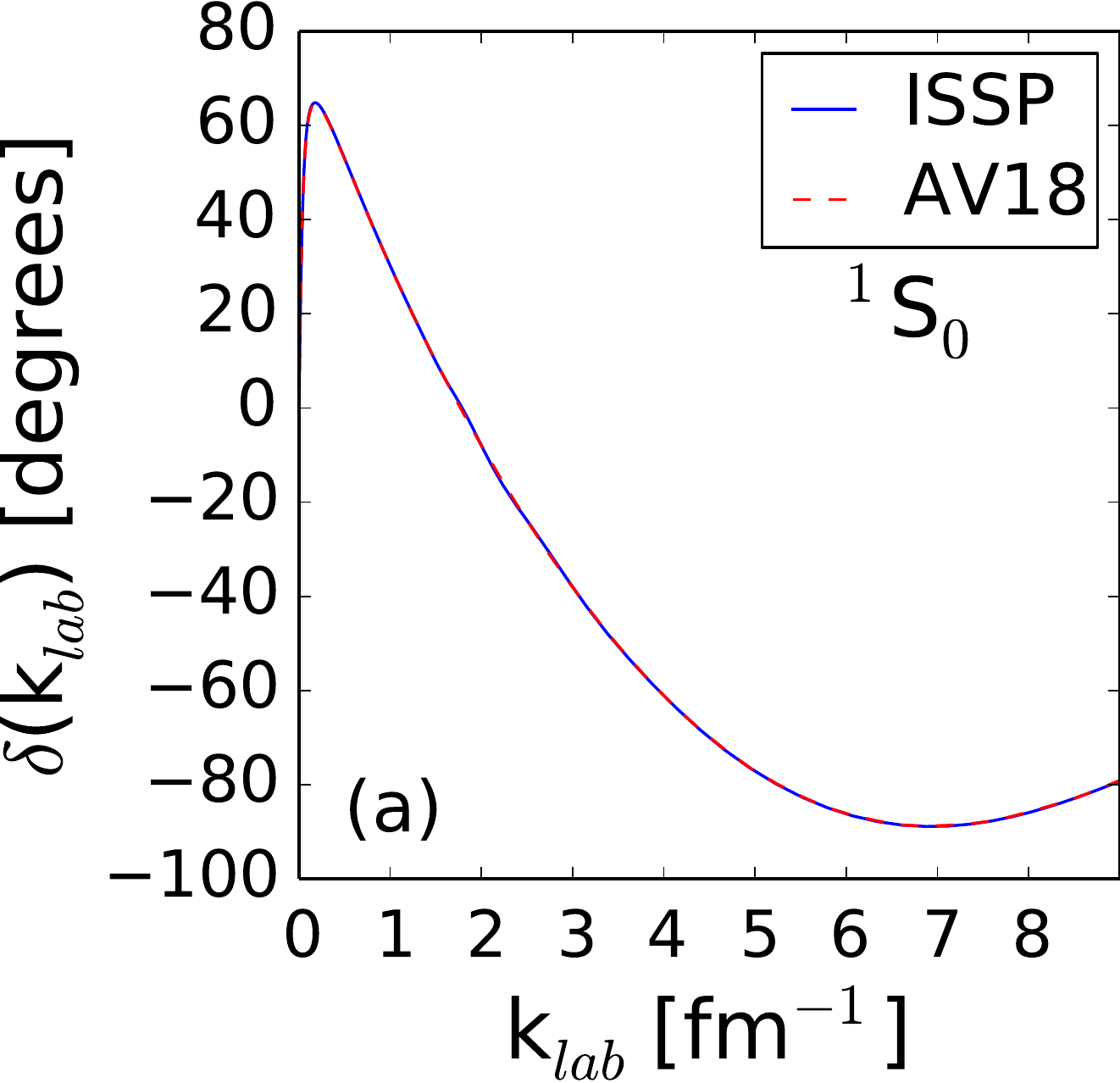}}
	\hfill
	{\includegraphics[width=.32\textwidth]{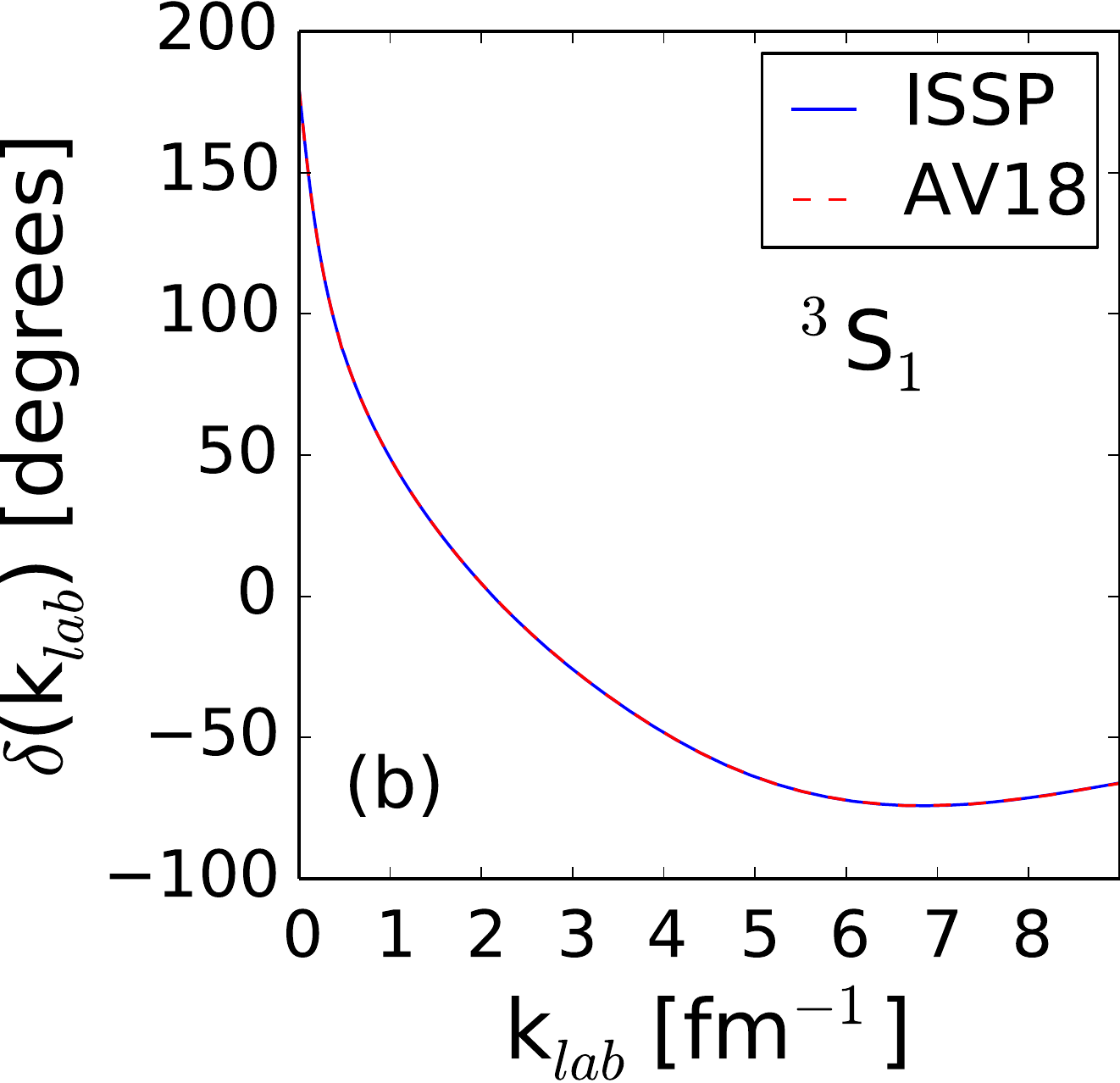}}
	\hfill
	{\includegraphics[width=.31\textwidth]{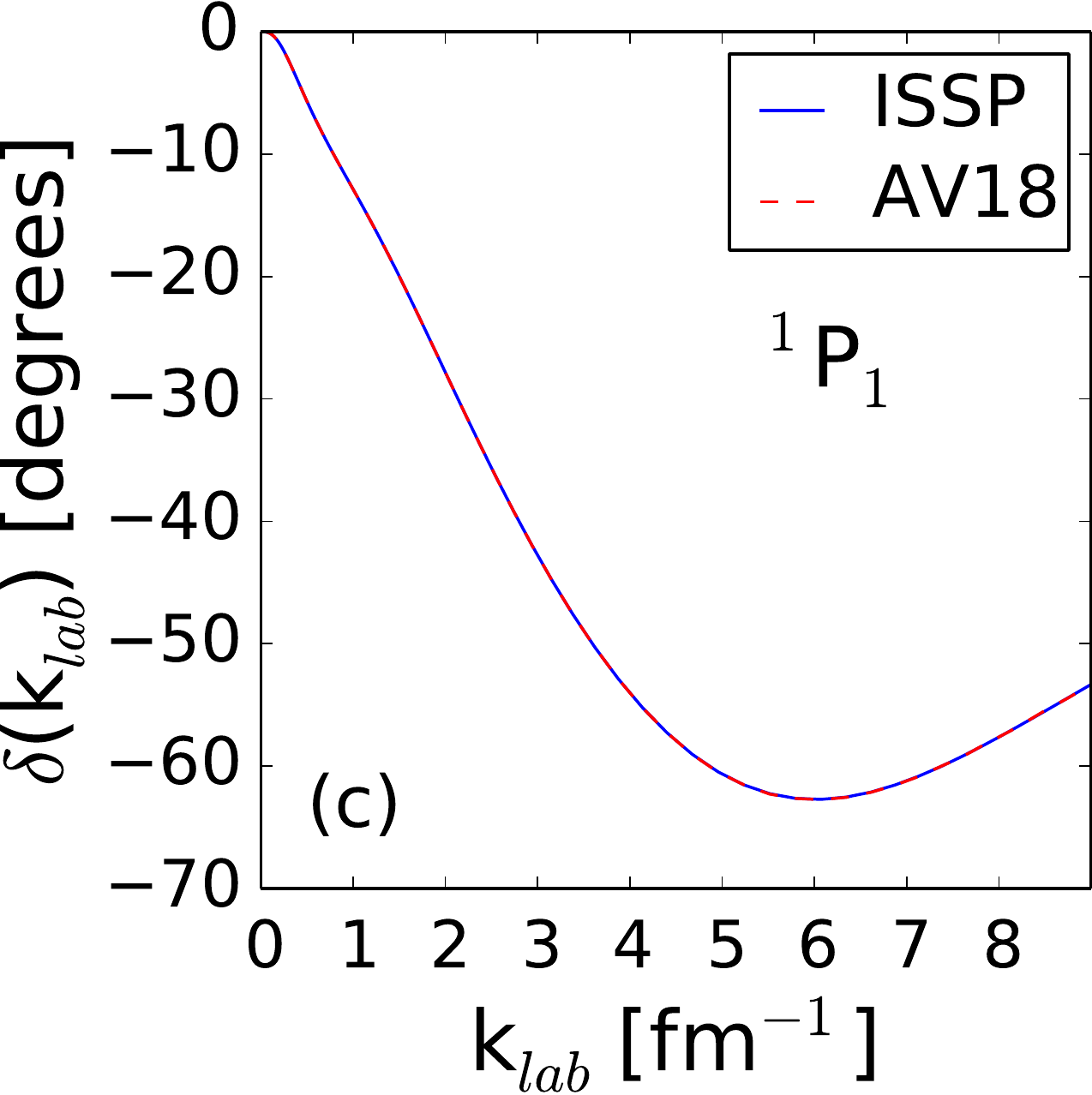}}
	\vspace*{-.1in}
	\caption{(Color online) Phase shifts using the AV18 potential and the ISSP up to high lab
	momentum $k_{\rm lab}$ in the (a) \oneSzero, (b) \threeSone, and
	(c) \onePone\ partial waves.
	\label{fig:ps_issp_av18_all}}
\end{figure*}

\begin{figure*}[htp!]
	{\includegraphics[width=.32\textwidth]{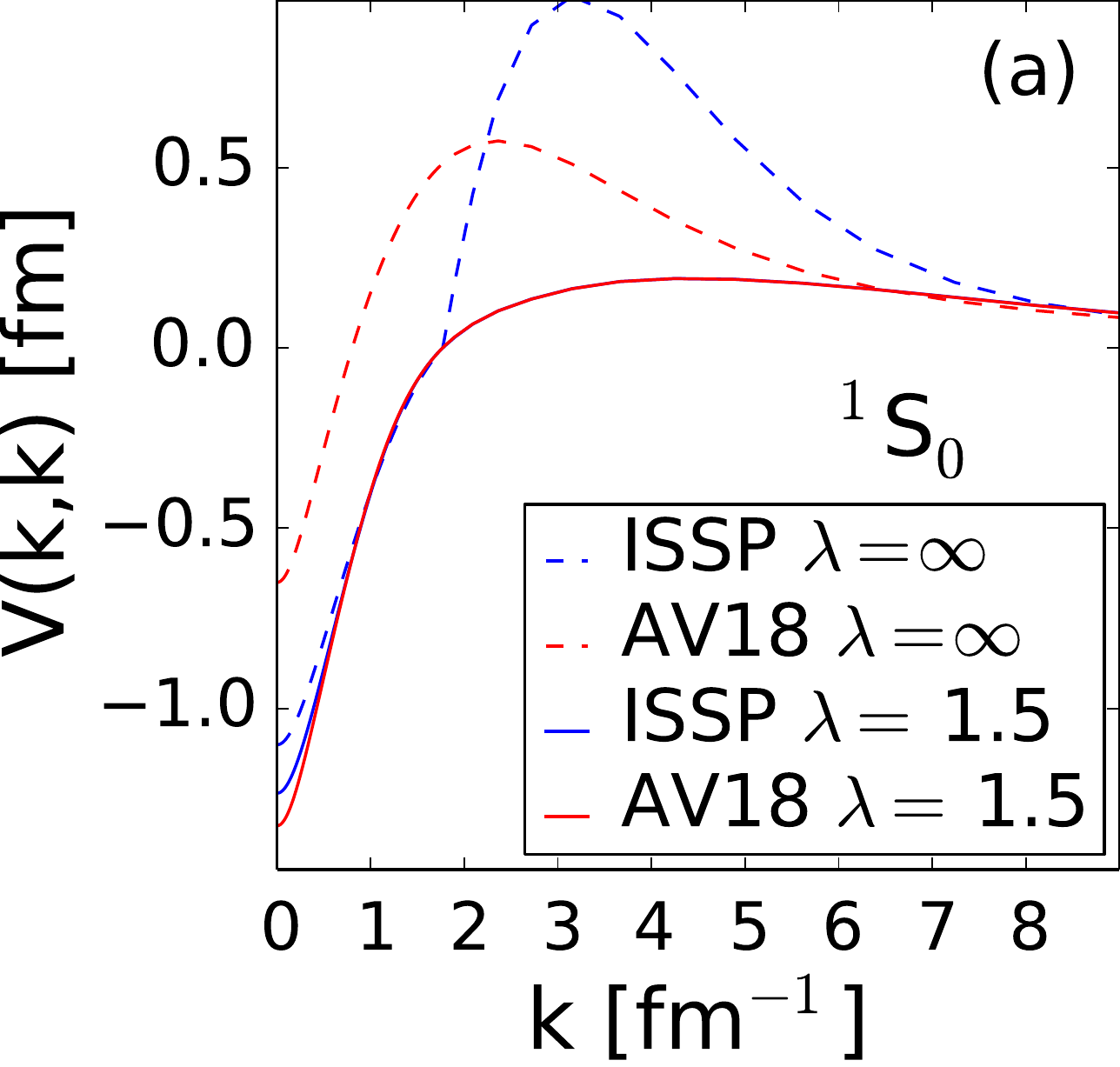}}
	\hfill
	{\includegraphics[width=.32\textwidth]{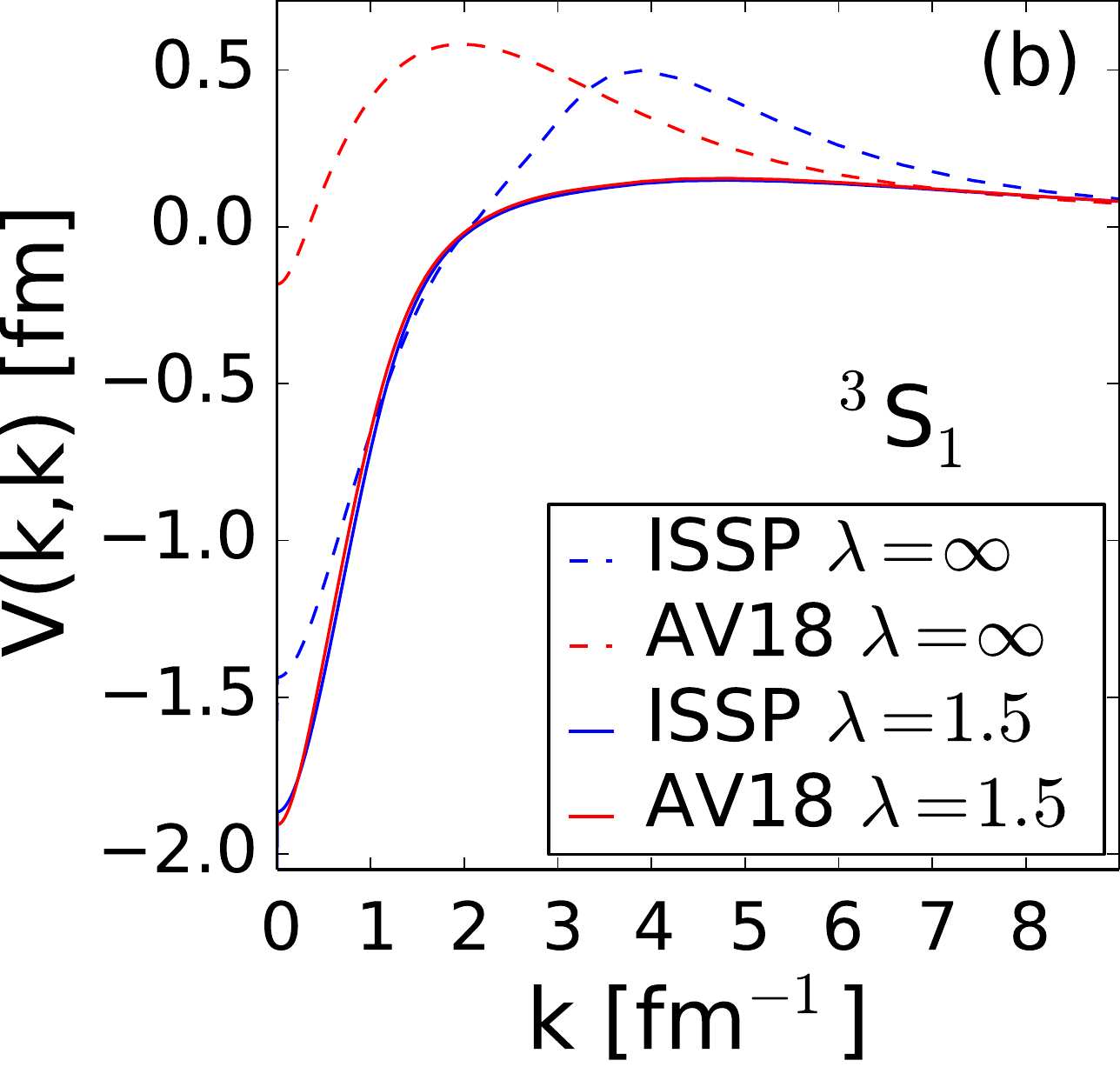}}
	\hfill
	{\includegraphics[width=.32\textwidth]{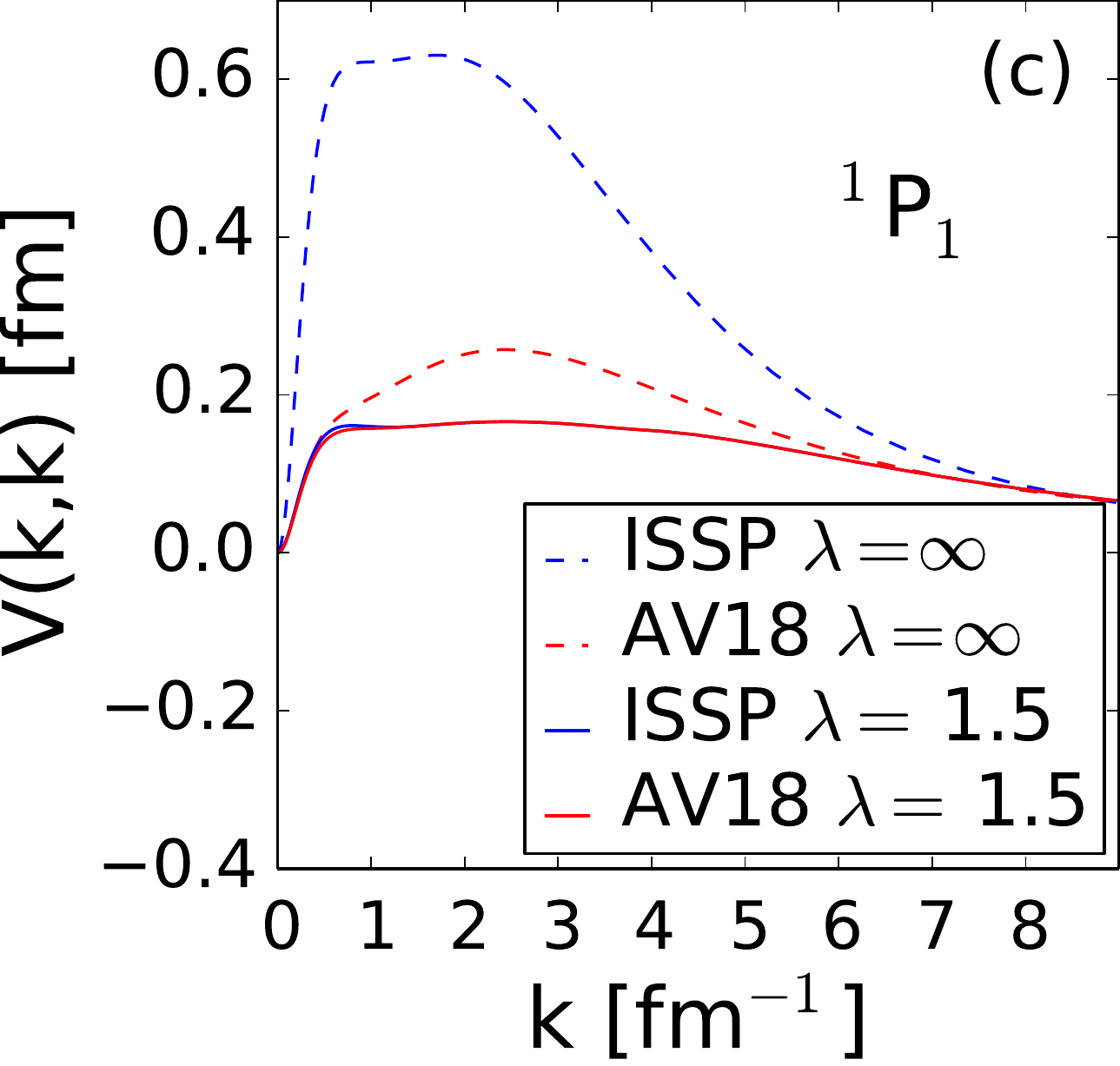}}
	\vspace*{-.1in}
	\caption{(Color online) Diagonal matrix elements of the AV18 potential and the ISSP up to high lab
	momentum $k_{\rm lab}$ in the (a) \oneSzero, (b) \threeSone, and
	(c) \onePone\ partial waves.  Cutoff $\lambda$ is in units of fm$^{-1}$.
	\label{fig:diagss_issp_av18_all}}
\end{figure*}



The method described thus far works directly for uncoupled channels, but for 
NN scattering we must also account for
coupled channels, where some further formalism is required.  
For this purpose, we use the Blatt-Beidenharn (BB) convention for phase shifts in the coupled channel~\cite{BBps,Kwong}.  (In the plots we employ the more typically used 
Stapp-$\overline{N}$ convention for the phase shifts~\cite{Nbarps}.)  
The BB convention can be summarized as:
	\begin{eqnarray}
		\mathbf{S}(k) &=& \mathbf{U}^{\dag}(k) \mathbf{\widehat{\Delta}}(k) \mathbf{U}(k) \;,\\
		\mathbf{\widehat{\Delta}}(k)&=& \begin{pmatrix} e^{2 i \delta_{0}(k)} & 0 \\ 
								  0 & e^{2 i \delta_{1}(k)}
			 			     \end{pmatrix} \;,\\
		\mathbf{U}(k)&=& \begin{pmatrix} \cos(\epsilon(k)) & \sin(\epsilon(k)) \\ 
								  -\sin(\epsilon(k)) & \cos(\epsilon(k))
			 			     \end{pmatrix} \;.
	\end{eqnarray}
Here, S(k) is the scattering matrix define in Ref.~\cite{BBps}, with k the momentum corresponding to
the interaction energy.
Then the inverse scattering potential can be written as:
\begin{equation}
		\mathbf{V}(k,k\rq{}) = 
		\mathbf{U}^{\dag}(k) \mathbf{\hat{V}}(k,k\rq{}) \mathbf{U}(k\rq{}) \;,
\end{equation}
where
\begin{equation}		
		\mathbf{\hat{V}}(k,k\rq{}) =  \begin{pmatrix} \hat{V}_{0}(k,k\rq{}) & 0 \\ 
								                        0 & \hat{V}_{1}(k,k\rq{})
			 			               \end{pmatrix} 
		\;.
\end{equation}
To proceed, one uses the inverse scattering method for uncoupled channels
to find $\hat{V}_{0}(k,k\rq{})$ from $\delta_{0}(k)$ and $\hat{V}_{1}(k,k\rq{})$ 
from $\delta_{1}(k)$.  
The complete potential is then found by a rotation by the mixing parameter, $\epsilon(k)$.  
With this complete separable inverse scattering formalism, we can now create a phase-equivalent potential at all energies in any given partial-wave channel.  

 
\subsection{Universality in separable inverse scattering potentials} \label{sec:universality}

We use phase shifts from \avpot\ to create the phase-equivalent ISSP.  
In Fig.~\ref{fig:ps_issp_av18_all}, we see that the elastic phase shifts are 
quantitatively reproduced well above the inelastic threshold.  
We choose \avpot\ specifically because it has phase shifts that extend to this high
energy, but any realistic potential could be used for starting phase shifts.  
(Note: for simplicity we treat the problem non-relativistically with only elastic scattering 
because we are interested in testing universality and low-energy effects, not to
have a realistic description of high-energy physics.)  
The ISSP\rq{}s from chiral potentials exhibit similar behavior, except that the internal 
cutoffs drive matrix elements and phase shifts to zero at high energies, which is
less useful for the present investigations.  
The accuracy of the ISSP in reproducing phase shifts can be further increased 
simply by using more grid points and increasing the maximum momentum if the phase 
shifts are nonzero above this momentum.  

Figure~\ref{fig:diagss_issp_av18_all} shows the diagonal matrix elements of \avpot\ 
and the ISSP for three different partial waves before and after SRG evolution.  
We observe that after SRG evolution to $\lambda=1.5$ fm$^{-1}$, universality in the 
diagonal matrix elements also extends to the full range of energies.  
In fact, the only discernible difference in the evolved potential diagonals is below 
the SRG cutoff.  
Above $\lambda$ the matrix elements in the region shown are completely collapsed to universal values.

\begin{figure}[th!]
	\includegraphics[width=3.2in]{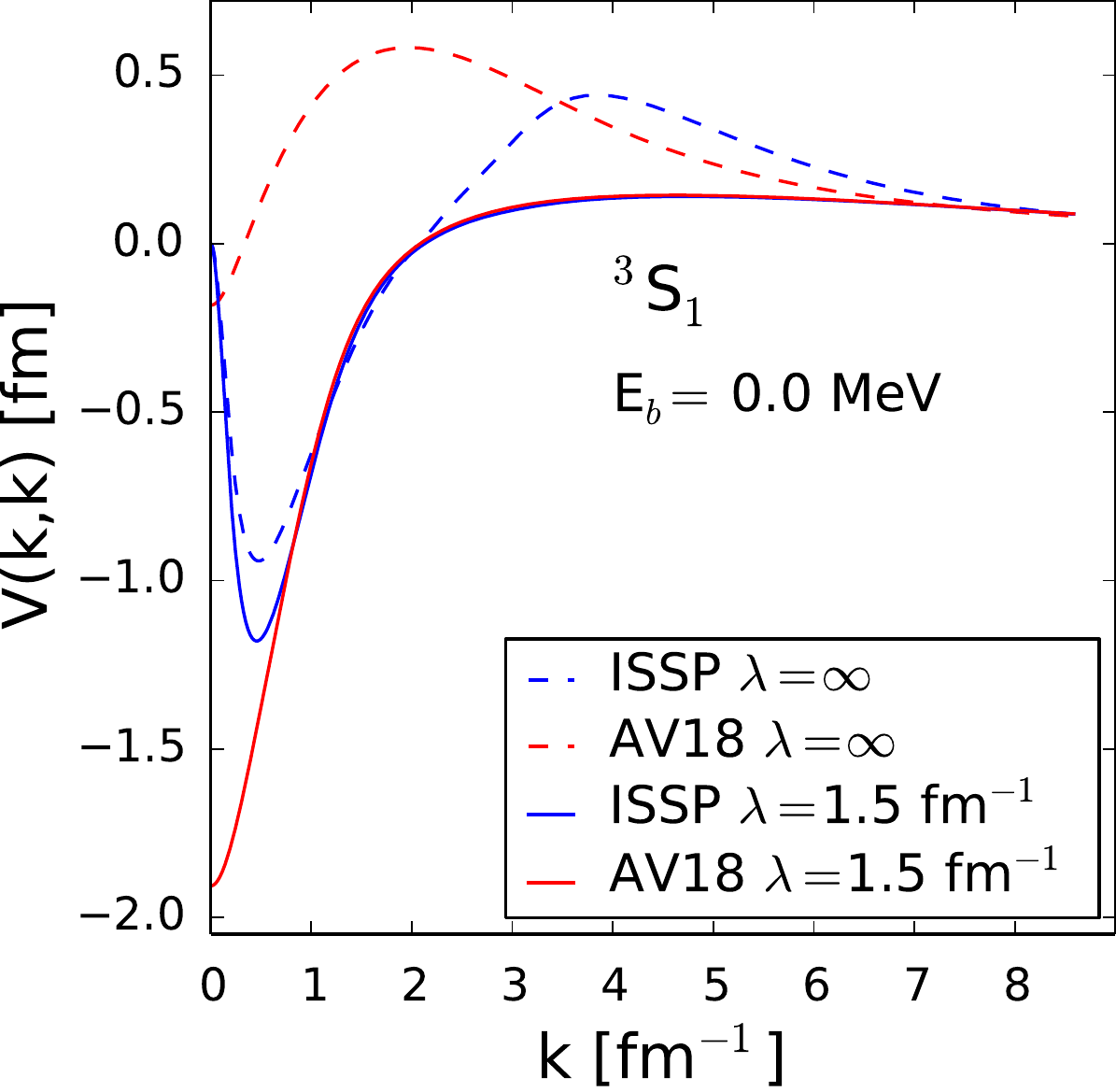}
	\caption{(Color online) Initial and evolved diagonal matrix elements in the $^{3}$S$_{1}$ channel
	for AV18 and an ISSP with a binding energy of 0\,MeV. 
	\label{fig:diagss_issp_av18_0_bd}}
\end{figure}

\begin{figure}[th!]
	\includegraphics[width=3.2in]{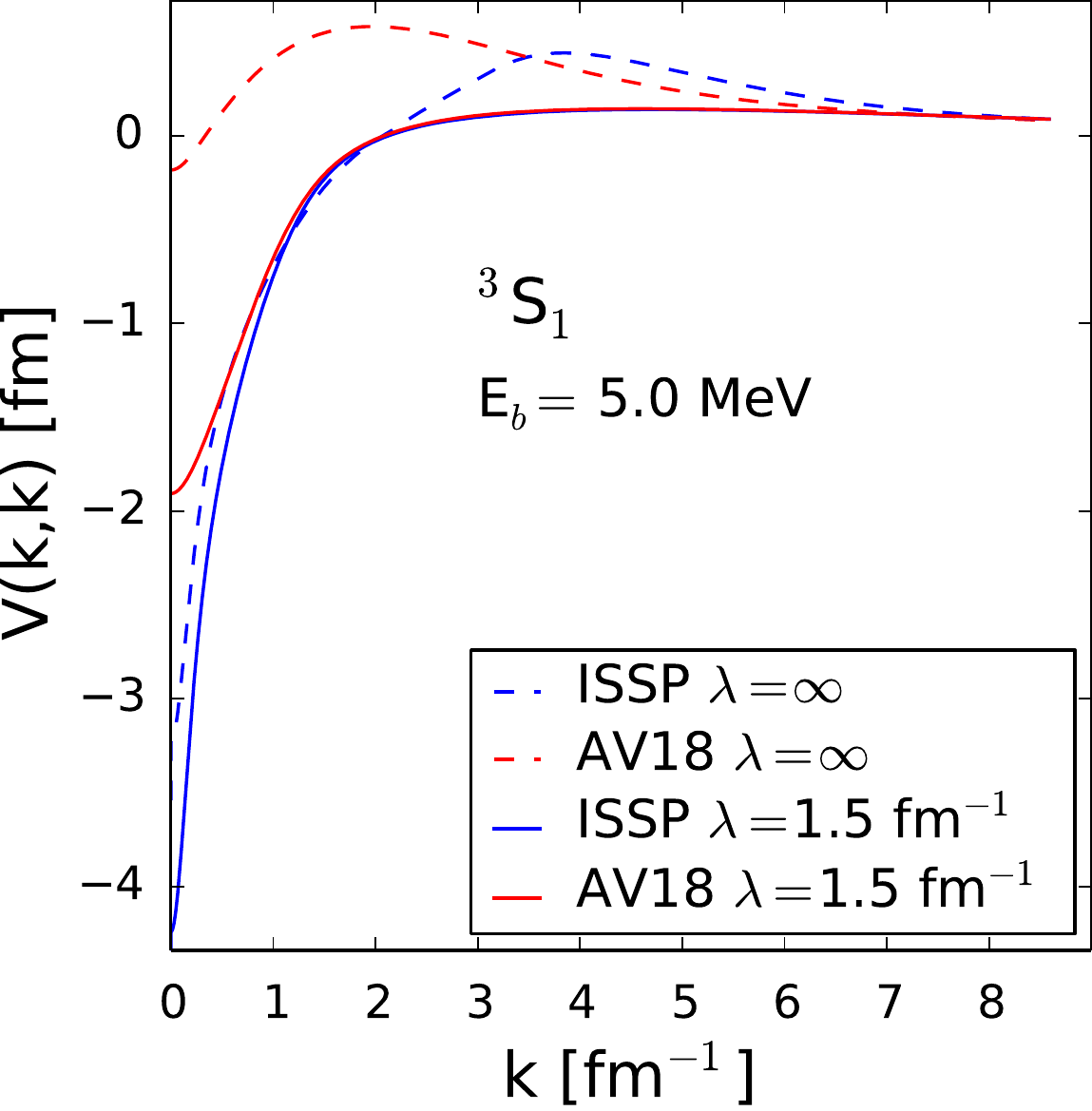}
	\caption{(Color online) Initial and evolved diagonal matrix elements in the $^{3}$S$_{1}$ channel
	for AV18 and an ISSP with a binding energy of $5$\,MeV. 
	\label{fig:diagss_issp_av18_5_bd}}
\end{figure}

Because the binding energy in the ISSP formalism is independently tuned from the phase shifts,
we can investigate in the deuteron
$^{3}$S$_{1}$-$^{3}$D$_{1}$ coupled channel how universality
in potential matrix elements is affected by differences in the bound-state energy.
Figures~\ref{fig:diagss_issp_av18_0_bd} and \ref{fig:diagss_issp_av18_5_bd} show the effect of
phase-equivalent potentials having the wrong binding energy.  
In Fig.~\ref{fig:diagss_issp_av18_0_bd}, the ISSP is created from the phase shifts of the 
\avpot\ potential in the deuteron channel, but with a binding energy of 0\,MeV
instead of 2.224\,MeV. 
It is evident that the effect on diagonal matrix elements is substantial.  
The low-energy matrix elements of the bare ISSP tend towards zero as the momentum decreases.  
As the potentials evolve, the diagonal matrix elements are driven to universal values except
that the ISSP is constrained by its binding energy to approach zero as momentum approaches zero.

A similar effect can be seen in Fig.~\ref{fig:diagss_issp_av18_5_bd} where instead of 0\,MeV as input binding energy, the ISSP is created with input binding energy of $5$\,MeV.  
The ISSP reproduces this energy better than 100\,eV.  
This potential is overbound and its lowest momentum matrix elements are forced lower 
than if it had the physical deuteron binding energy.  
Again, the higher momentum matrix elements flow towards a universal form because of phase equivalence.  
Together these plots show that phase equivalence is not the only prerequisite for universality in the diagonal potential matrix elements, but a correct binding energy is also necessary.  
(That is, we need S-matrix equivalence for negative energies as well.)
This may account for the small deviations in the potentials at lowest momenta in 
Fig.~\ref{fig:diag_modern_inf}.  The $^{3}$D$_{1}$ partial wave plots of the corresponding ISSP potentials with different binding energies are indistinguishable.  
This effect only appears in the \threeSone\ potentials.  
It is possible that a virtual bound state in the \oneSzero\ partial wave has a similar effect 
on the evolved low-momentum potential matrix elements, but the ISSP cannot tune virtual bound states and residues in the same way it accommodates bound states, thus we do not investigate this point further.

\begin{figure*}[tp!]
	{\includegraphics[width=.44\textwidth]{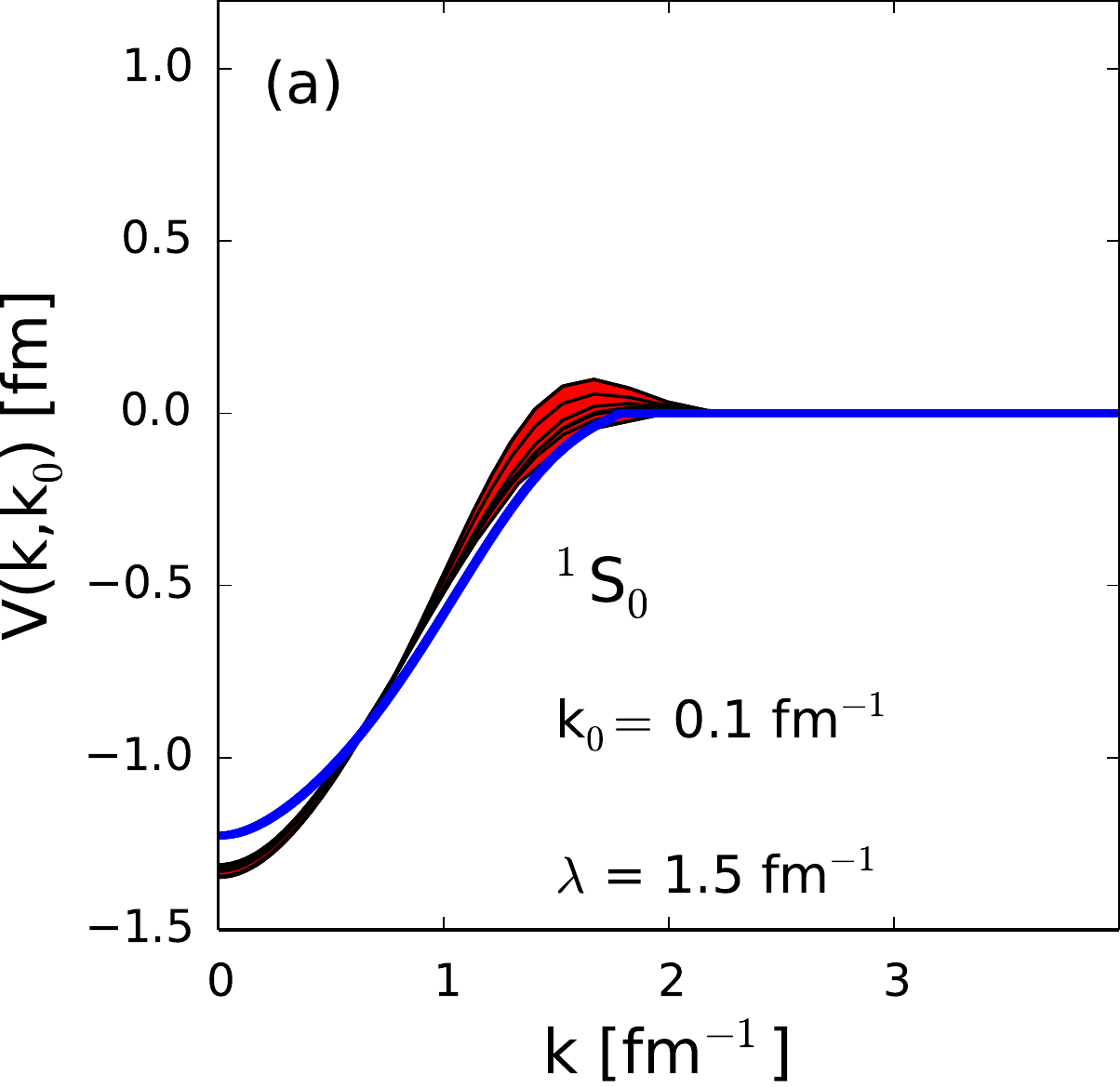}}
	\hspace*{.5in}
	{\includegraphics[width=.44\textwidth]{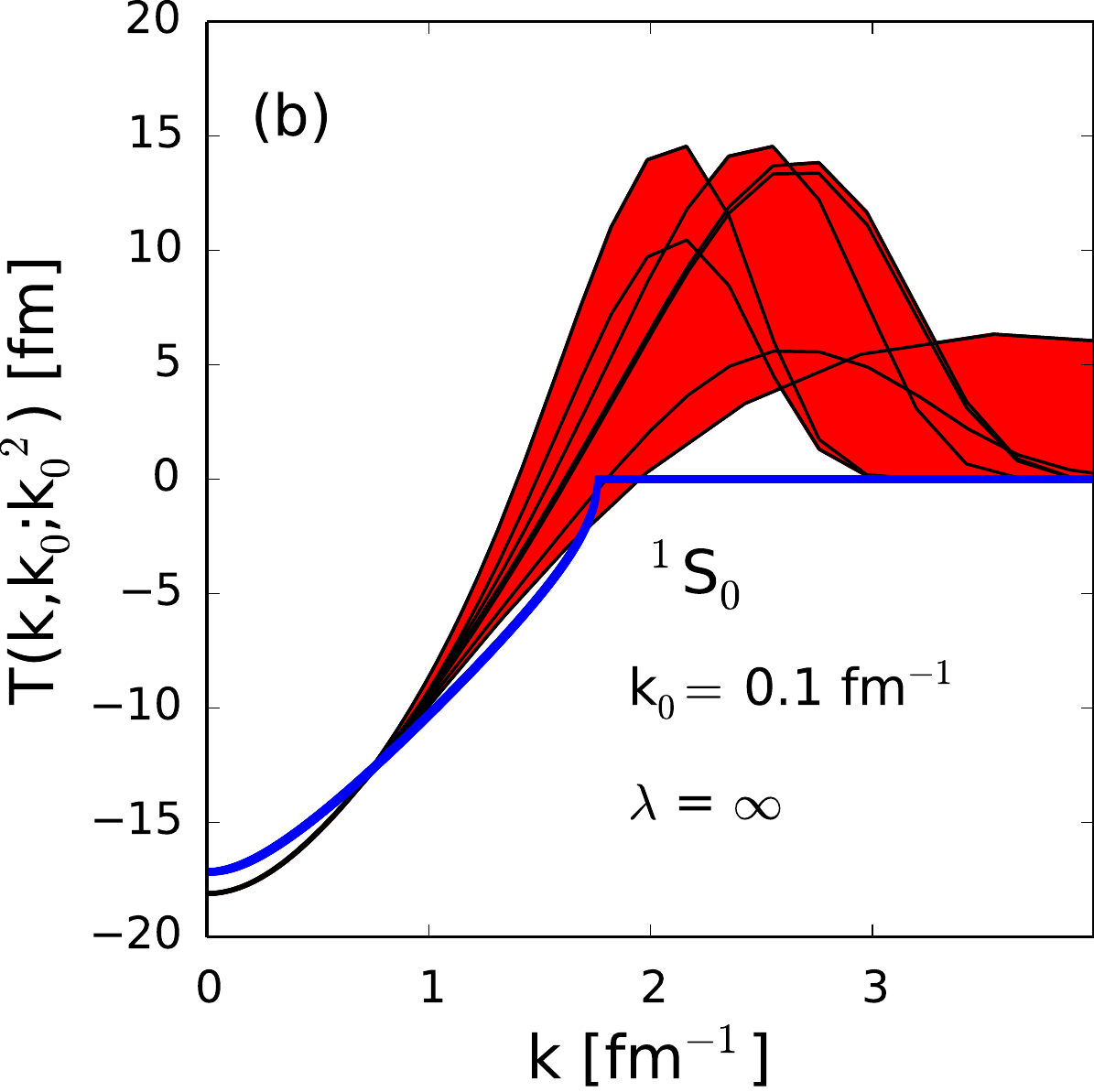}}
	\vspace*{-.1in}
	\caption{(Color online) (a) Off-diagonal SRG evolved potential matrix elements
	$V(k,k_0)$ with $k_0 = 0.1\infm$ and (b) unevolved half-on-shell T matrices
	$T(k,k_0;k_0^2)$.  
	In both figures, the thick line is from the ISSP while the bands are various realistic
	potentials. 
	\label{fig:isspno}}
\end{figure*}


Next we turn to off-diagonal matrix elements.
Figure~\ref{fig:isspno} shows the potential matrix elements $V(k_0,k)$ for $k_0 = 0.1\infm$
as a function of $k$ for the ISSP and all of the realistic potentials evolved to $\lambda=1.5\infm$.  
We can see that although these off-diagonal cuts for the modern potentials agree at 
$\lambda=1.5\infm$, the ISSP matrix elements do not.  
By using a diagonalizing SRG transformation (that is, $G_s = T$), the off-diagonal potential
matrix elements are exponentially suppressed.  
Because of this, it appears that the ISSP \emph{approaches} a universal form, but unlike the
realistic potentials, there is no finite $\lambda$ at which the ISSP collapses to universal 
form.  
Figure~\ref{fig:isspno} shows low-energy half-on-shell (HOS) T matrices from each of the unevolved realistic potentials and the ISSP.  
We observe that the realistic potentials, which will evolve to a universal form, have 
essentially the same low-momentum, low-energy HOS T-matrix elements, 
while the ISSP does not.  
This is consistent with carrying over to the SRG the suggestion from Ref.~\cite{vlowkuniv} that HOS T-matrix equivalence
is required for off-diagonal universality in $\vlowk$ RG-evolved matrix elements, 
much like phase shift equivalence is required for universality of diagonals.  
We only show the \oneSzero\ partial waves, but the same pattern holds for all partial waves.  
Clearly, matching observables is not enough to produce fully universal potentials 
after evolution, and in the next section we will examine if matching observables and also including
the same explicit one-pion exchange potential will be enough for potentials to evolve to a low-energy
universal form.

\begin{figure*}[bhtp!]
	{\includegraphics[width=.44\textwidth]{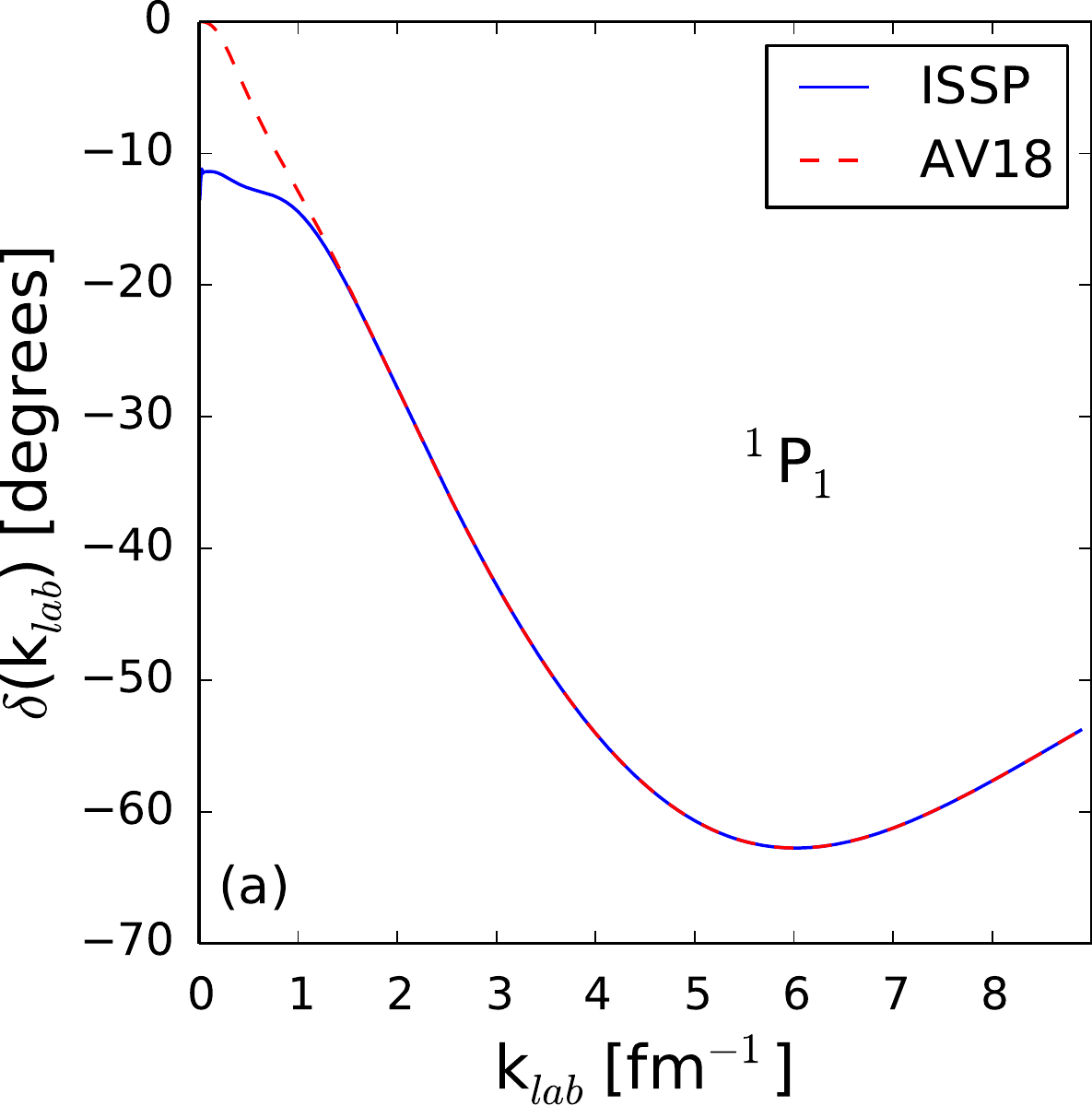}}
	\hspace*{.5in}
	{\includegraphics[width=.44\textwidth]{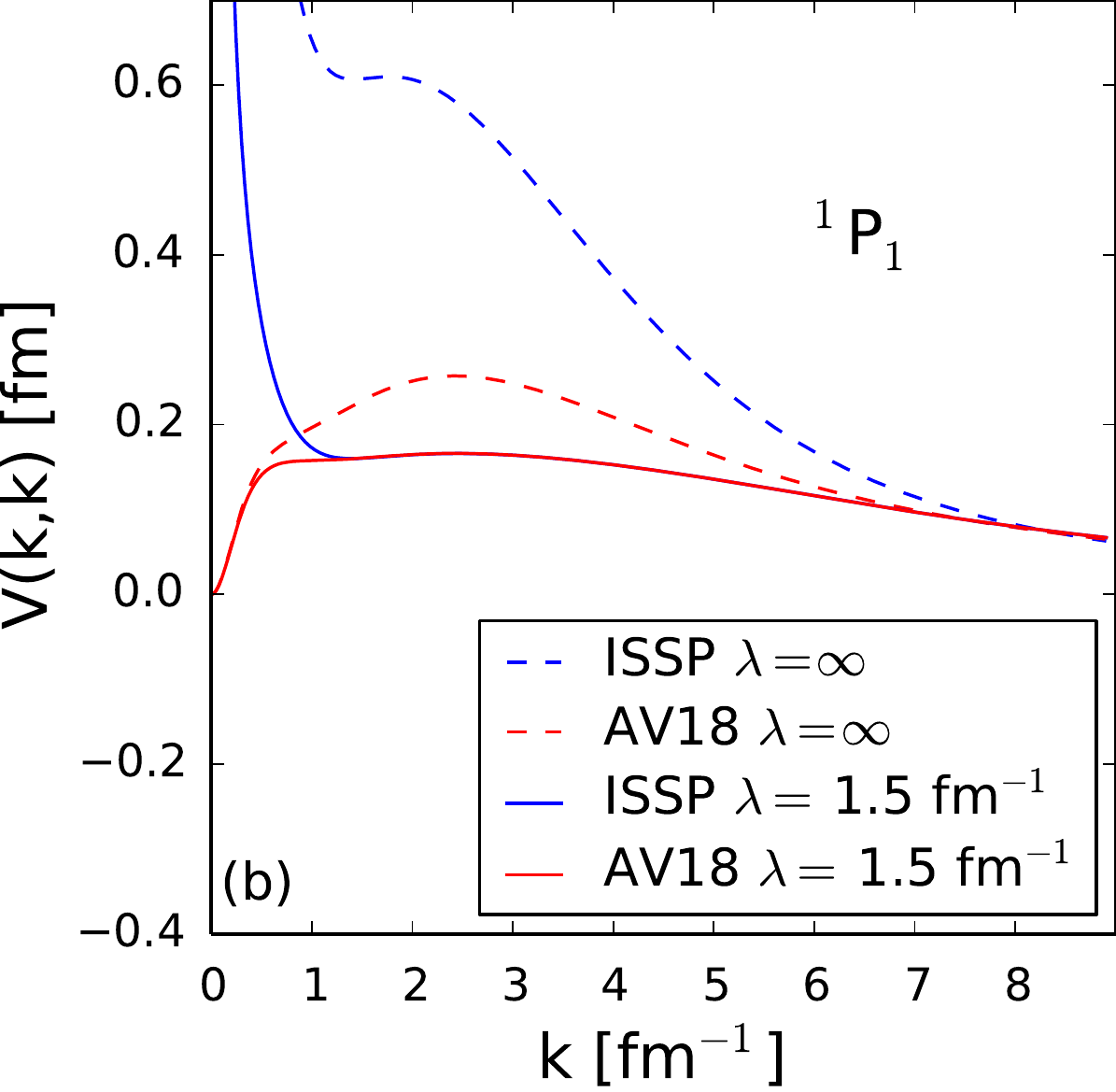}}
	\vspace*{-.1in}
	\caption{(Color online) Low-energy phase difference effects on universality  (a) phase shifts, (b) diagonals of potentials\label{fig:altered_low}}
\end{figure*}

\begin{figure*}[htp!]
	{\includegraphics[width=.44\textwidth]{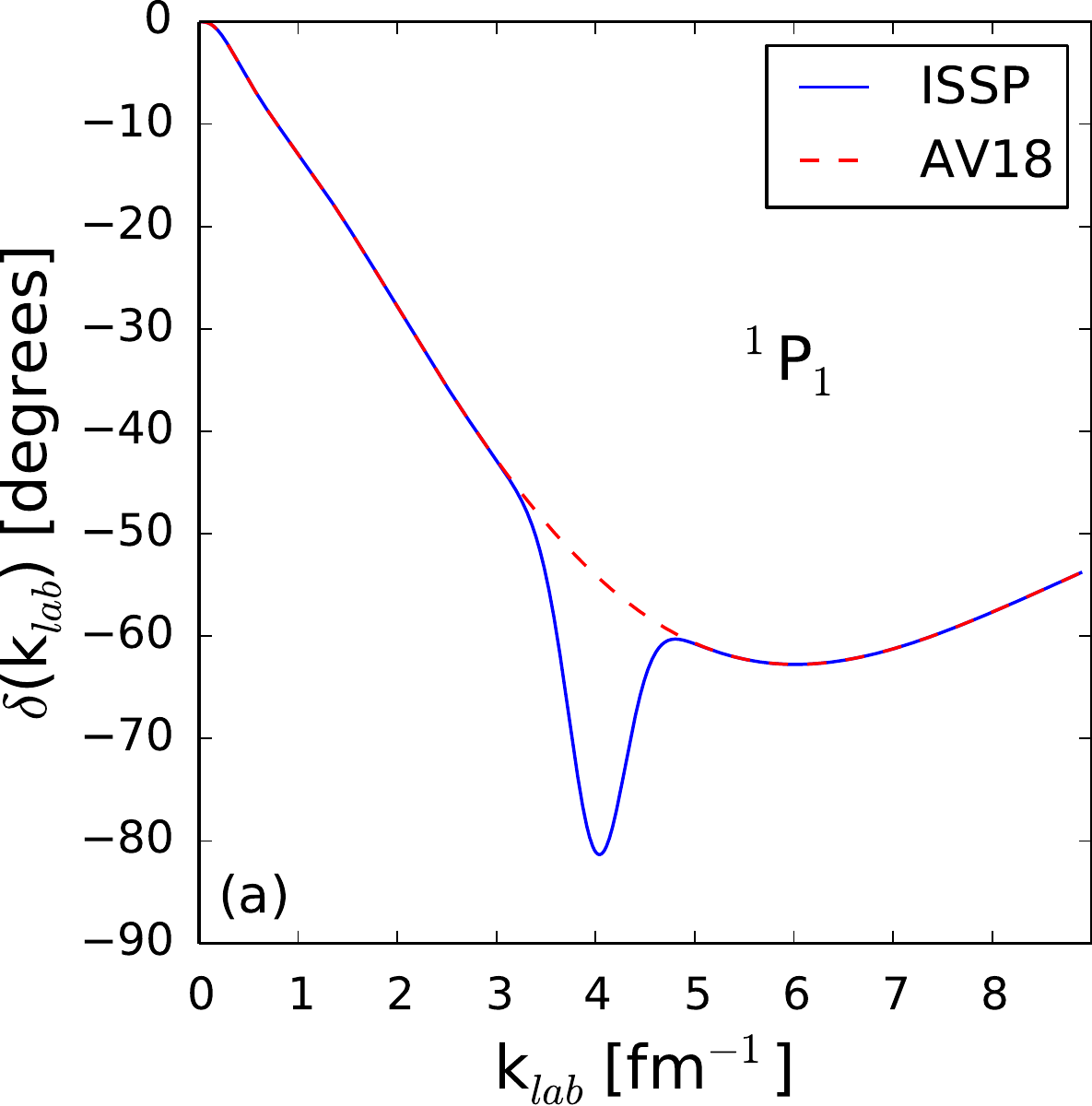}}
	\hspace*{.5in}
	{\includegraphics[width=.44\textwidth]{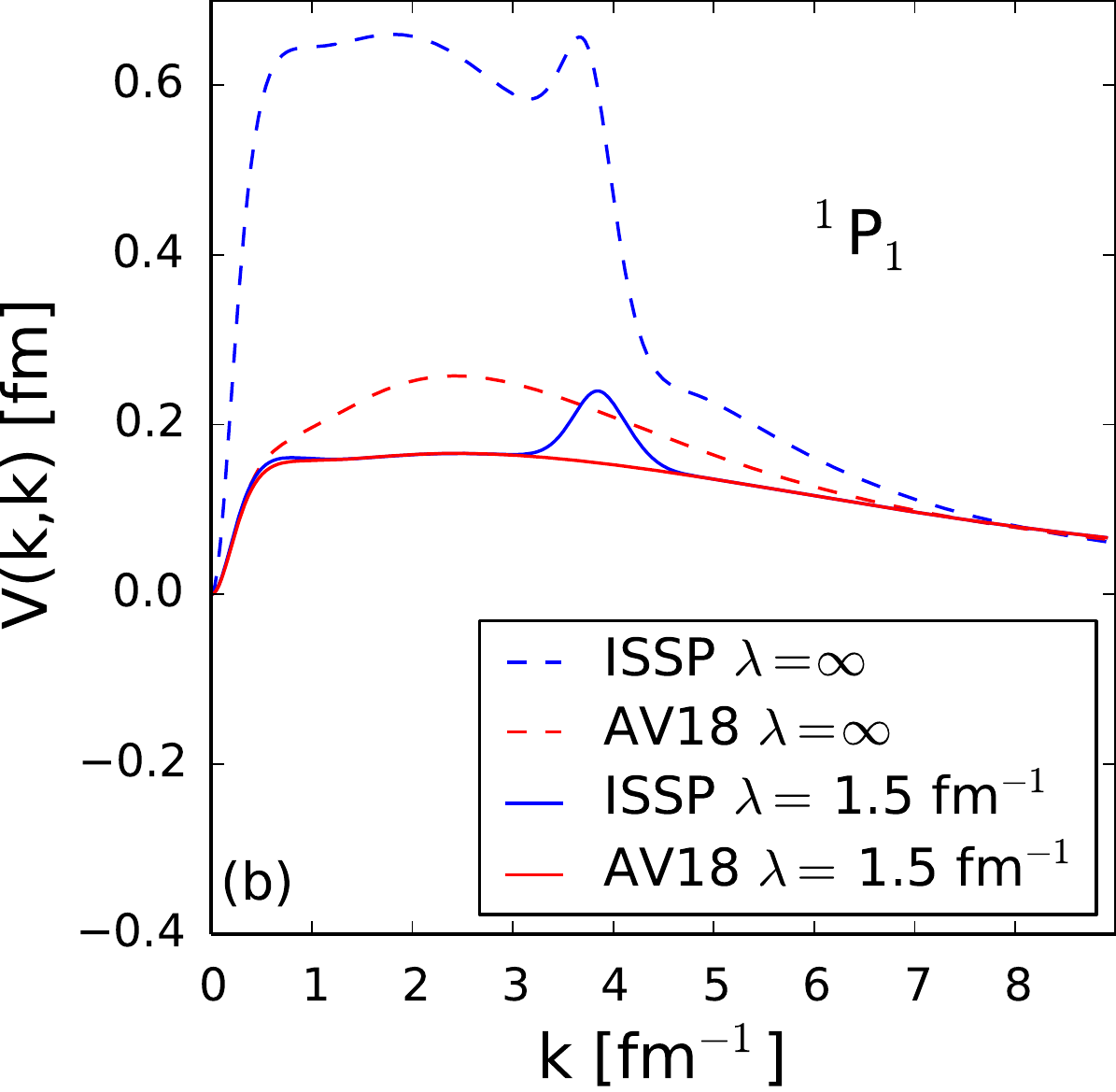}}
	\vspace*{-.1in}
	\caption{(Color online) Intermediate-energy phase difference effects on universality.  (a) phase shifts, (b) diagonals of potentials\label{fig:altered_bump}}
\end{figure*}



\begin{figure}[tbhp!]
	\includegraphics[width=3.2in]{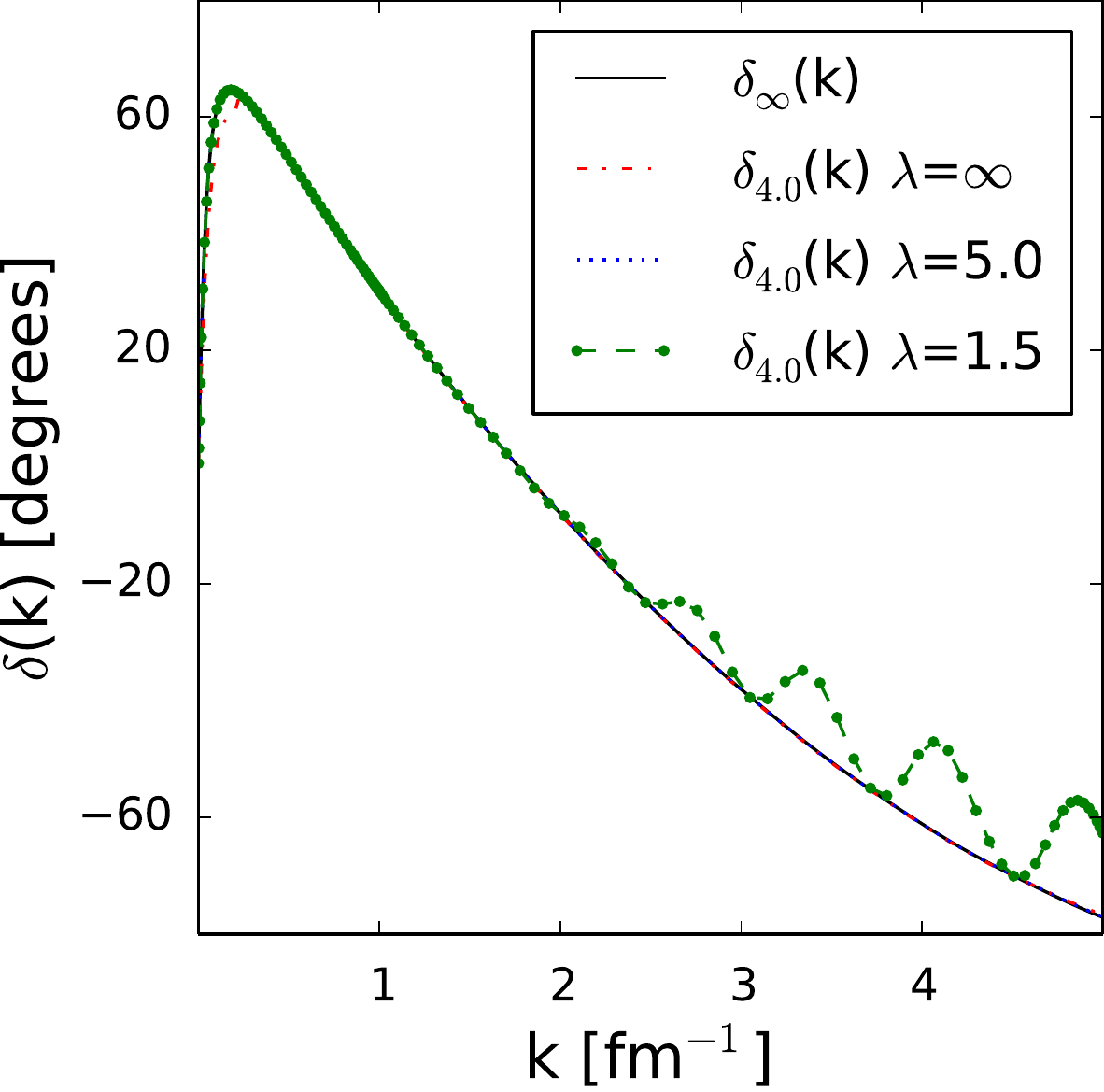}
	\caption{(Color online) Phase shifts of Argonne $v_{18}$ potential and truncated phase shifts of evolved potentials with $\Lambda$ = 4.0 fm$^{-1}$.  Cutoff $\lambda$ is in units of fm$^{-1}$.
	\label{fig:localps40}}
\end{figure}

\begin{figure}[tbhp!]
	\includegraphics[width=3.2in]{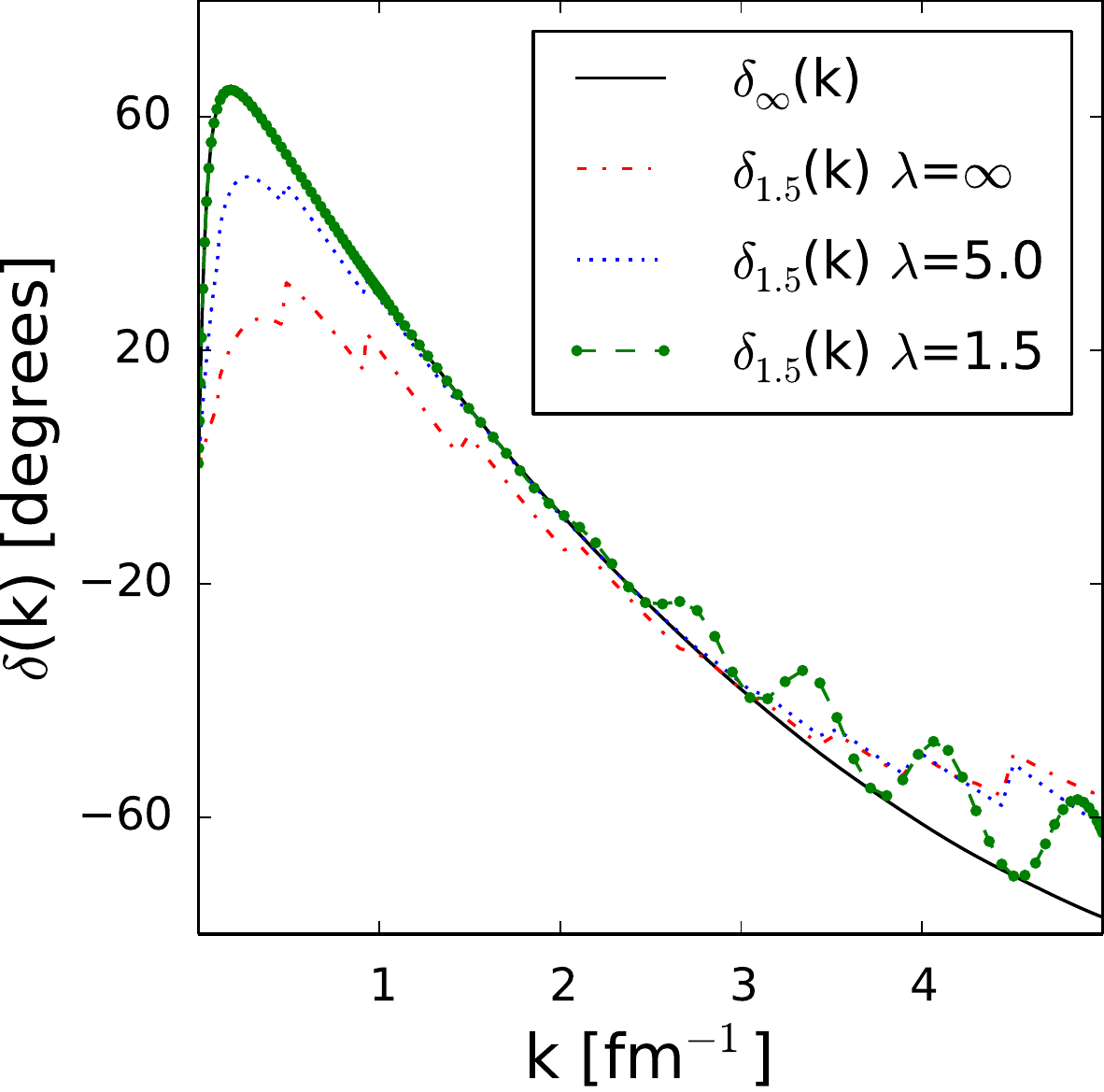}
	\caption{(Color online) Phase shifts of Argonne $v_{18}$ potential and truncated phase shifts of evolved potentials with $\Lambda$ = 1.5 fm$^{-1}$.  Cutoff $\lambda$ is in units of fm$^{-1}$.
    \label{fig:localps15}}
\end{figure}

As a further test, we created ISSP\rq{}s using altered phase shifts in \emph{localized} regions
of energy to see if the flow to universal diagonal matrix elements is disturbed
only locally.
We use the \onePone\ channel for clarity.  
Figure~\ref{fig:altered_low}(a) shows the \onePone\ phase shifts for \avpot\ and 
for an ISSP that is phase equivalent \emph{except} for a Gaussian bump that we
impose by hand at low energy.
In Fig.~\ref{fig:altered_low}(b) we see that the potentials evolve to the same
diagonal values everywhere but at low energy.
Another potential was constructed by creating low- and high-energy regions of phase equivalence,
and imposing a Gaussian bump (around $k_{\rm lab}=4.0\infm$) to create a difference in the intermediate energy phase shifts, see Fig.~\ref{fig:altered_bump}(a).  
In Fig.~\ref{fig:altered_bump}(b) the evolution to common diagonal values again
works everywhere except near where the phase shifts disagree. 

We conclude from these figures (and other tests not shown) 
that the SRG evolved diagonal potential matrix elements are 
altered only in a region localized near the altered phase shifts.  
This suggests that an SRG softened potential is \emph{locally decoupled} such that 
the integral in the Lippmann-Schwinger (LS) equation for the on-shell T matrix can be truncated as:
\begin{eqnarray}
	T_{l}(k,k;k^2) = V_{l}(k,k)  
	   &+& \frac{2}{\pi} P\!\int_{k-\Lambda}^{k+\Lambda} dp\, p^{2} 
   \nonumber \\ & &  \null \times
	\frac{V_{l}(k,p)T_{l}(p,k;k^2)}{k^{2}-p^{2}}
	\;,
	\label{localLS}
\end{eqnarray}
where the lower limit of the integral is taken to be zero if $k - \Lambda < 0$.
In Eq.~\eqref{localLS}, $\Lambda$ represents the local decoupling scale, which we will set to 
SRG $\lambda$.
(In fact $\lambda$ appears to be a conservative upper bound for $\Lambda$ to quantitatively reproduce phase shifts.)

Figure~\ref{fig:localps40} shows phase shifts calculated from
Eq.~\eqref{localLS} with $\Lambda = 4.0\infm$ 
in the \oneSzero\ channel for the Argonne $v_{18}$ potential evolved to three different SRG 
$\lambda$'s.  These are compared to the actual phase shifts of the unevolved potential.  
We see that with this large value of $\Lambda$, the truncated phase shifts
for even the unevolved potential are largely reproduced and  
the low-momentum phase shifts from evolved potentials are indistinguishable from
the actual phase shifts.  
(The periodicity at high momentum for $\lambda = 1.5\infm$ is a numerical
grid artifact.) 
In Fig.~\ref{fig:localps15} we more severely truncate the integral in the LS equation 
to $\Lambda = 1.5\infm$.  
We see clearly that the potential evolved to $\lambda = 4.0\infm$ is not 
decoupled enough to 
reproduce the original phase shifts, but the potential evolved to $\lambda = 1.5\infm$ 
has phase shifts identical to the previous plot.  
This suggests that evolution with $T$ does locally decouple energy scales.


\begin{figure*}[htp!]
	{\includegraphics[width=.44\textwidth]{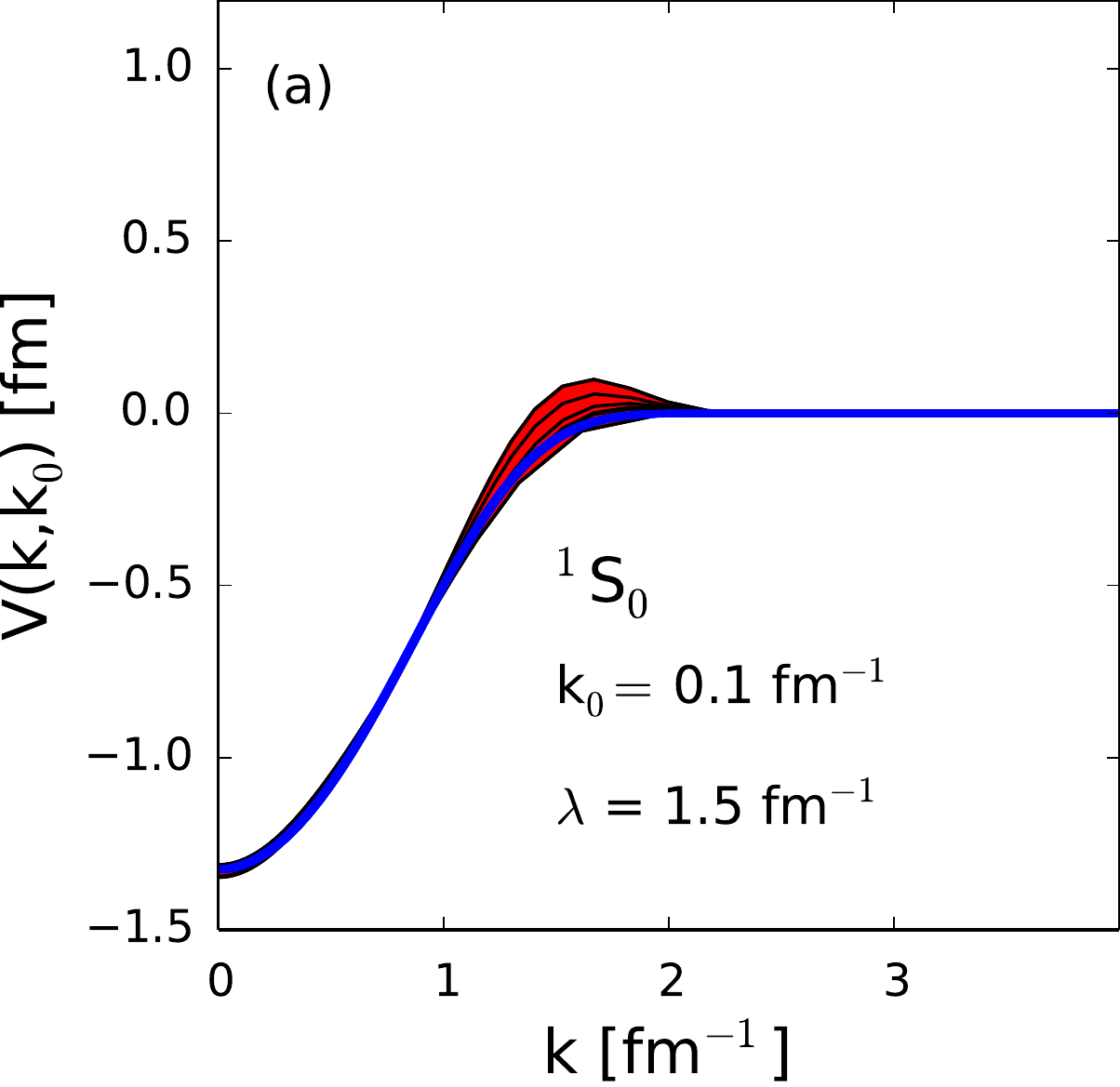}}
	\hspace*{.5in}
	{\includegraphics[width=.44\textwidth]{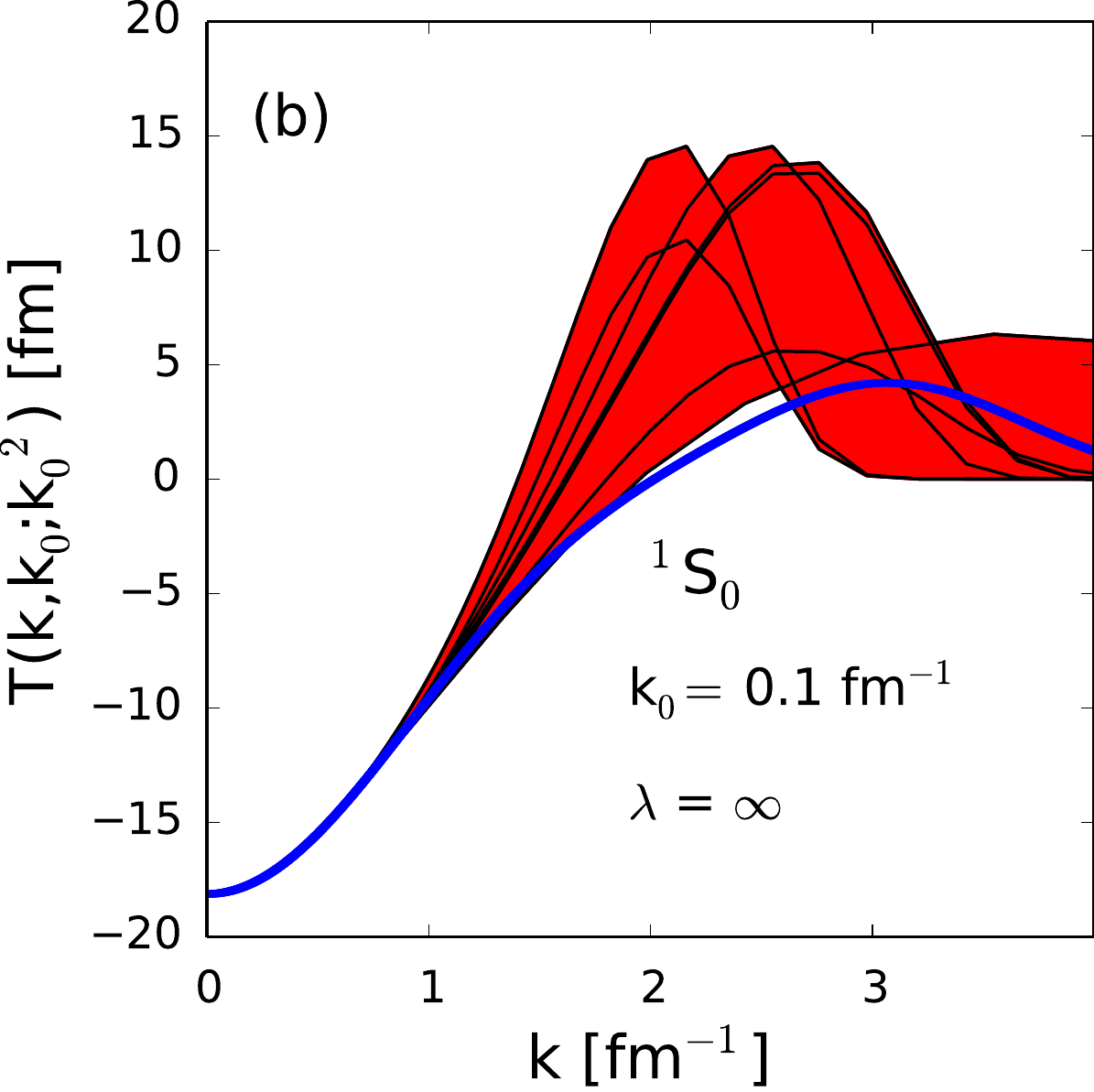}}
	\vspace*{-.1in}
	\caption{(Color online) (a) Off-diagonal SRG evolved potential matrix elements. (b) Unevolved half-on-shell T matrices.  In both figures, the thick line is the $\delta$-shell plus OPE potential 
	while the bands are from realistic modern potentials. 
	\label{fig:dsuniv}}
\end{figure*}

\begin{figure*}[htp!]
	{\includegraphics[width=.32\textwidth]{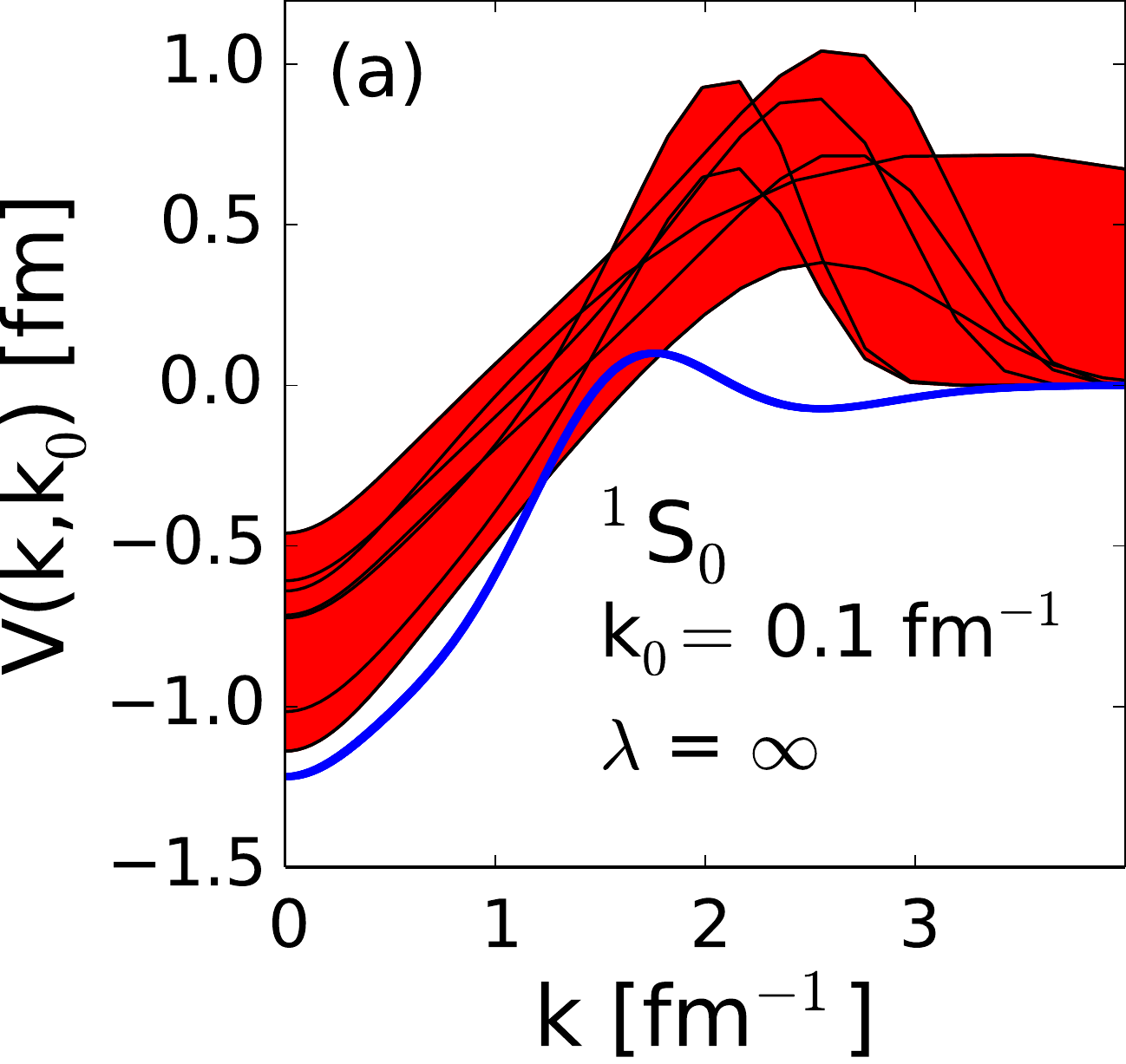}}
	\hfill
	{\includegraphics[width=.32\textwidth]{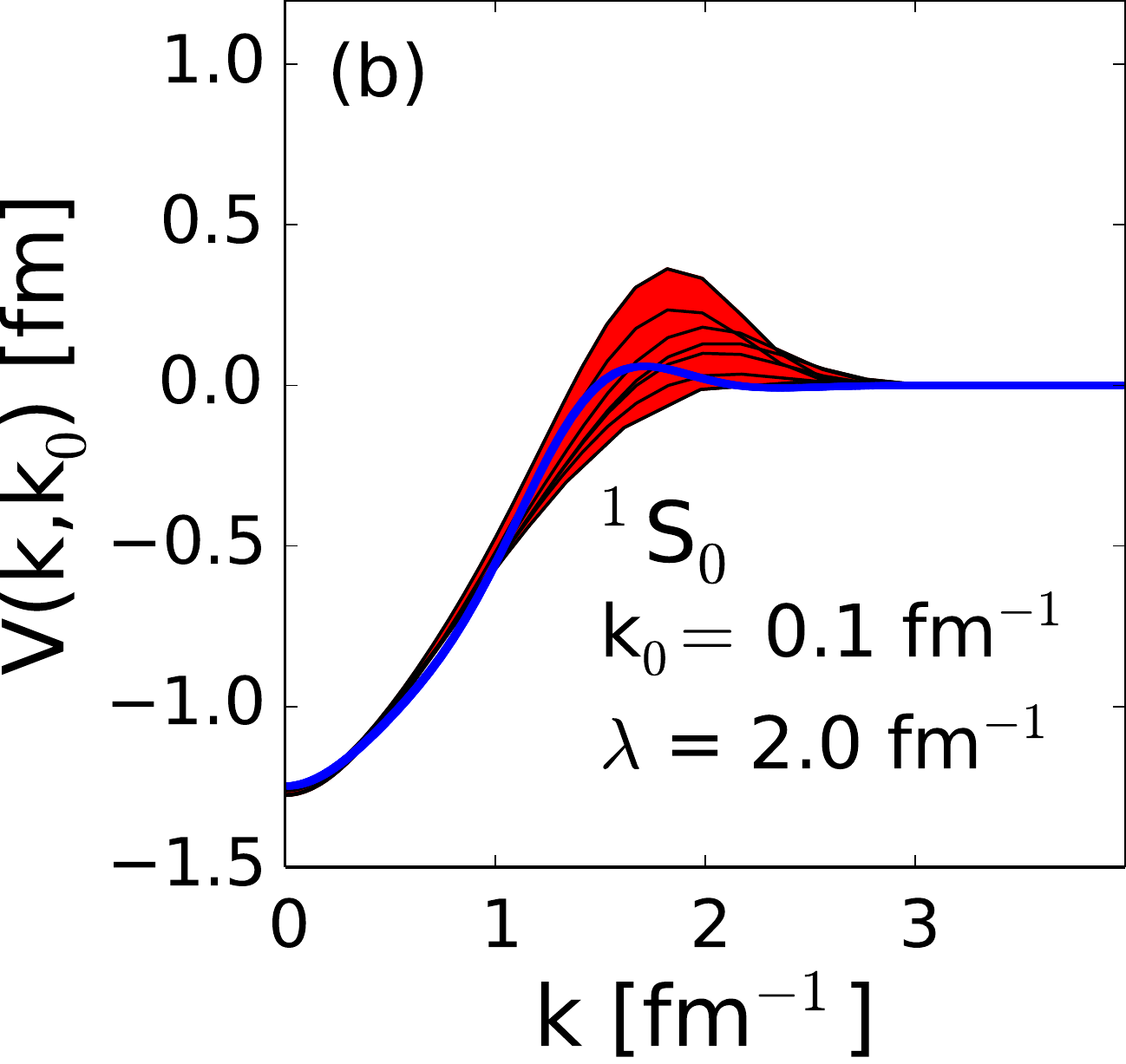}}
	\hfill
	{\includegraphics[width=.32\textwidth]{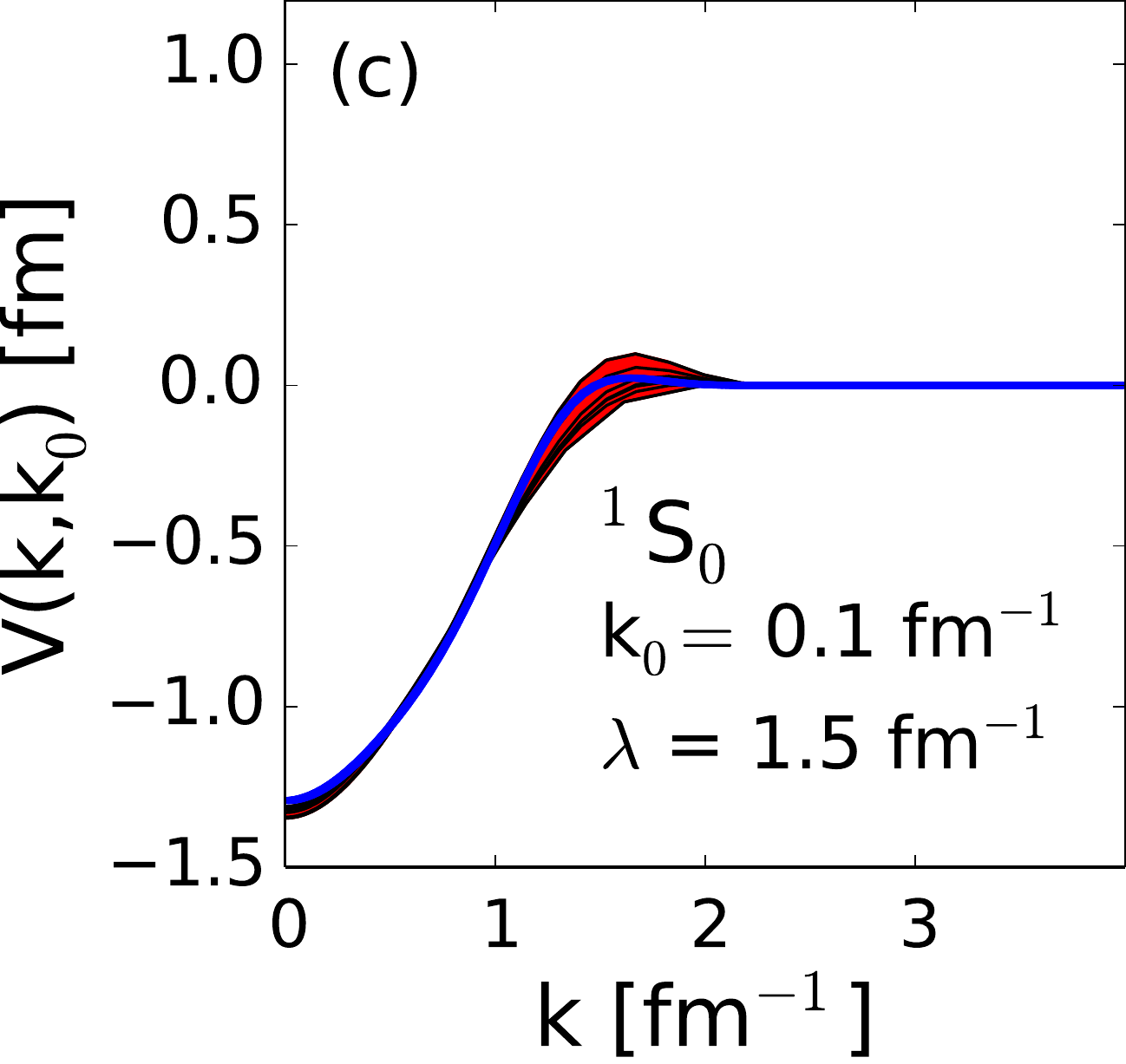}}
	\vspace*{-.1in}
	\caption{(Color online) Off-diagonal matrix elements of (a) chiral N$^3$LO potentials and the JISP16 potential  
	in the \oneSzero\
  channel and the same potential evolved by the SRG to (b) $\lambda = 2.0\infm$ and (c) $\lambda = 1.5\infm$.
	\label{fig:1s0_jisp_all}}
\end{figure*}




\section{OPE plus $\delta$-shell} \label{sec:OPE}

Here we further test the suggestion that explicit treatment of the longest-ranged physics is a 
requirement for potentials to evolve to a universal form~\cite{vlowkuniv}. 
In particular, we develop a simple test potential that is (approximately)
phase equivalent in the same
momentum regions as the realistic potentials but also has the same explicit long-range forces.  
We use the model from Navarro P\'erez \textit{et al}.\ that combines the one-pion exchange (OPE) potential with a sum of $N$ $\delta$-shell potentials~\cite{Arriola,Perez:2013jpa} in each
partial wave:  
	\begin{equation}
		V_{l}(r)= V_{l}^{\rm OPE}(r) + \sum_{i=1}^{N} g^l_{i} \delta(r-r_{i}) \;.
	\end{equation}
The explicit form of the OPE potential can be found in Ref.~\cite{OPE}.  
We choose the $\{r_{i}\}$ as short-range lengths (under $2\infm$), 
and fit the $\{g^l_{i}\}$ to match low-momentum phase shifts.  
We choose a different regulator than Ref.~\cite{Arriola,Perez:2013jpa}, instead regulating the
potential in momentum representation with a separable form factor:
	\begin{equation}
		f_{\rm reg}(k,k')= e^{-({k}/{\Lambda})^{4}}e^{-({k'}/{\Lambda})^{4}} \;,
	\end{equation}
for which we choose $\Lambda = 3$ fm$^{-1}$.  
We now have a potential with explicit long-range pion terms and adjustable short range terms, 
which is phase equivalent at low momentum to the realistic potentials.


\subsection{Universality in OPE plus $\delta$-shell}

We can see from Fig.~\ref{fig:dsuniv} that the OPE plus $\delta$-shell 
off-diagonal potential elements evolve to the same universal form as the modern realistic potentials.  
Also, Fig.~\ref{fig:dsuniv} shows the corresponding unevolved HOS T matrices.  
We see that the OPE plus $\delta$-shell potential has the same low-energy low-momentum 
HOS T matrix and shows a corresponding low-momentum universality in off-diagonal matrix 
elements.  
This behavior is not unique to the \oneSzero\ partial wave, but appears for all partial waves.  
This simple potential explicitly contains only the longest range OPE potential and has very simple short-range terms, but it collapses to the same universal low-momentum potential 
after SRG evolution.  
Combined with the ISSP results, this is strong evidence that the same explicit inclusion of
the longest-range contributions to the potential, which is reflected in
low-energy HOS T-matrix equivalence, is required for collapse to a universal form.  

\subsection{JISP potential}  \label{subsec:jisp}

In principle, a good test of our observations about universality is the JISP16 potential,
which is a realistic potential
constructed using the $J$-matrix version of inverse scattering theory~\cite{Shirokov:2003kk,Shirokov:2005bk}.
Because there is no explicit incorporation of a pion-exchange tail in the functional
form of the potential, we might expect the Hamiltonian to exhibit non-universal evolution
with the SRG for off-diagonal matrix elements.
In fact, the unevolved JISP potential is already soft and changes only slightly under
SRG evolution.  
But as shown in Fig.~\ref{fig:1s0_jisp_all} in the \oneSzero\ channel for a set of off-diagonal
matrix elements (and true for the diagonal and other partial waves), JISP16 is already close
to the universal form reached by the chiral N$^3$LO potentials.  There are still differences,
but they are small.  However the JISP HOS T matrix is also close to the others (perhaps as
the result of additional adjustments of the potential using the freedom of the inverse
scattering framework ~\cite{Shirokov:2005bk}), so there is no inconsistency with our general conclusions.


\section{Conclusion}  \label{sec:conclusion}

Modern realistic  two-nucleon potentials exhibit a
flow to universal potential matrix elements under the similarity
RG. High and low momenta are decoupled in this universal matrix,
allowing us to truncate the matrix and drastically simplify 
low-energy bound state and reaction calculations. Any initial interaction
that yields this universal matrix after SRG evolution is equally effective. This is of little
practical importance for the two-nucleon potential, but it could be extremely
useful if many-nucleon potentials display this same type of universality. Producing
accurate realistic many-nucleon potentials is extremely difficult. Our results 
suggest that any convenient potential that includes long-range pion exchange 
interactions can be used to produce universal many-nucleon interactions
when evolved with an SRG transformation.
 
Our study of universality for two-body potentials
yields the following observations:
\begin{itemize}
 \item Inverse scattering separable potentials, with no explicit consideration 
of long-range pion exchange, exhibit a universal collapse
of diagonal matrix elements  after evolution
in regions of phase equivalence. 

\item
If an intermediate region of phase inequivalence
is imposed, the collapse does not occur in this region, but still occurs in every region of phase equivalence.
This implies that SRG softened potentials are actually \emph{locally}
decoupled in energy/momentum. 

 \item
An incorrect binding energy has a strong effect on the lowest potential 
matrix elements and will prevent flow towards a universal form.  

\item
  While phase equivalence and correct binding energies (i.e., S-matrix equivalence) are apparently
requirements for universality in two-body potential matrix elements,  
the ISSP example shows that these are not sufficient to guarantee a potential 
that will flow to the same off-diagonal values as conventional realistic
potentials. 

\item
However, a potential that reproduces low-energy observables and 
contains explicit long-range (OPE) terms does flow to universal form, which
is consistent with observations made for $\vlowk$ evolution in
Ref.~\cite{vlowkuniv}.

\item
To the extent that low-energy HOS T-matrix equivalence indicates long-range
equivalence of potentials, it signals off-diagonal universality in evolved
potential matrix elements.

\smallskip

\item
For universality to appear, the SRG decoupling parameter must be sufficiently low that potential 
matrix elements in the low-momentum region of HOS T-matrix equivalence are decoupled 
from high-momentum matrix elements.

\end{itemize}

\noindent
These considerations address the onset of universality for the two-body part
of the inter-nucleon potential but
for a complete discussion we have to consider
the full many-body Hamiltonian.  It is well established that the evolution 
to smaller values of $\lambda$ induces many-body forces of increasing 
importance~\cite{bognerfurnstahlschwenk,Furnstahl:2013oba} and the SRG transformations
will only be approximately unitary if they are omitted.
This entails 
a lower limit to the region of universality in practical applications.
In future work 
we will test whether our observations of universality for two-nucleon interactions
carry over to three-body forces and seek a practical operator classification
procedure that will allow the full power of the RG to be applied to 
nuclear problems.

\vspace*{.1in}

\begin{acknowledgments}
For helpful comments and discussions we would like to thank S.~Bogner, K.~Hebeler, 
H.~Hergert, S.~More, and K.~Wendt.  
We also thank K.~Wendt for the use of several Python modules.  
This work was supported in part by the National Science Foundation under Grant 
Nos.~PHY--1002478 and PHY--1306250 and the U.S.\ Department of Energy under Grant No.~DE-SC0008533 (SciDAC-3/NUCLEI project).
\end{acknowledgments}


\bibliographystyle{apsrev4-1}
\bibliography{ISPuniversality3}

\end{document}